# Imaging Orbital-selective Quasiparticles in the Hund's Metal State of FeSe

A. Kostin[1,2*], P.O. Sprau[1,2*], A. Kreisel[3*], Yi Xue Chong[1,2], A.E. Böhmer[4,5], P.C. Canfield[4,6], P.J. Hirschfeld[7], B.M. Andersen[8] and J.C. Séamus Davis[1,2,9†]

1. Department of Physics, Cornell University, Ithaca, NY 14853, USA.
2. CMPMS Department, Brookhaven National Laboratory, Upton, NY 11973, USA.
3. Institut für Theoretische Physik, Universität Leipzig, D-04103 Leipzig, Germany
4. Ames Laboratory, U.S. Department of Energy, Ames, IA 50011, USA
5. Karlsruhe Institute of Technology, 76131 Karlsruhe, Germany
6. Department of Physics and Astronomy, Iowa State University, Ames, IA 50011, USA.
7. Department of Physics, University of Florida, Gainesville, Florida 32611, USA
8. Niels Bohr Institute., University of Copenhagen, Juliane Maries Vej 30, DK 2100 Copenhagen, Denmark
9. School of Physics and Astronomy, University of St. Andrews, Fife KY16 9SS, Scotland.

\* Contributed equally to this project.
† Corresponding author jcseamusdavis@gmail.com

**Strong electronic correlations, emerging from the parent Mott insulator phase, are key to copper-based high temperature superconductivity (HTS). By contrast, the parent phase of iron-based HTS is never a correlated insulator. But this distinction may be deceptive because Fe has five active *d*-orbitals while Cu has only one. In theory, such orbital multiplicity can generate a Hund's Metal state, in which alignment of the Fe spins suppresses inter-orbital fluctuations producing orbitally selective strong correlations. The spectral weights $Z_m$ of quasiparticles associated with different Fe orbitals $m$ should then be radically different. Here we use quasiparticle scattering interference resolved by orbital content to explore these predictions in FeSe. Signatures of strong, orbitally selective differences of quasiparticle $Z_m$ appear on all detectable bands over a wide energy range. Further, the quasiparticle interference amplitudes reveal that $Z_{xy} < Z_{xz} \ll Z_{yz}$, consistent with earlier orbital-selective Cooper pairing studies. Thus, orbital-selective strong correlations dominate the parent state of iron-based HTS in FeSe.**

The undoped phase proximate to superconductivity in copper-based materials is a strong Mott insulator[1,2], while that proximate to iron-based superconductivity is generally of a metallic nature[3,4]. This has motivated a perception that the mechanisms of high temperature superconductivity must be quite different in these two canonical materials classes and, moreover, that strong electronic correlations are not indispensable to HTS. Importantly, however, the electronic structure of the iron-based materials can still be governed by intense electronic correlations if an orbital-selective Hund's metal state exists[5-13]. This remarkable situation was discovered in theoretical studies of the multi-orbital Hubbard model (see the introduction to Supplementary Information) which typically consider the intra-orbital Hubbard energy U, the inter-orbital Coulomb interaction energy U' (=U-2J for spin-rotational symmetry), and the inter-orbital Hund's interaction energy J between spins. For a range of strong J, dynamical mean field theory[2] (DMFT) predicts that inter-orbital charge fluctuations are greatly suppressed, leading to an orbital decoupling of the strong correlations[6-13]. The striking consequence is that strongly correlated and thus low coherence states associated with one orbital are predicted to coexist with coherent delocalized quasiparticle states associated with the other.

In theory, Hund's metals occur in a region of intermediate to strong U and of strong J[5-13]. They are dominated by orbital-selective correlations, with the result that quasiparticle weights $Z_m$ associated with different orbitals $m$ diminish differently with increasing J or U. The quasiparticle weight $Z$ is given by $Z(\mathbf{k}) = (1 - \partial Re\Sigma(\mathbf{k},\omega)/\partial \omega |_{\omega=0})^{-1}$, where $\Sigma(\mathbf{k},\omega) = Re\Sigma(\mathbf{k},\omega) + iIm\Sigma(\mathbf{k},\omega)$ is the self-energy of a quasiparticle state $|\mathbf{k}\rangle$ with momentum $\hbar\mathbf{k}$ that is subject to strong electron-electron interactions. The quasiparticle weight on band j in $\mathbf{k}$-space, $Z_j(\mathbf{k})$, can be connected to the quasiparticle weight in orbital space $Z_m$ via the matrix elements of a unitary transformation. Multi-band Hubbard theories also exhibit orbital-selective quasiparticle (OSQP) phenomenology in which $Z_m$ evolves differently for each orbital $m$[2,7,9-10,14-16]. Moreover, when orbital degeneracies are lifted, for example by crystal field splitting in an orthorhombic/nematic phase, this further suppresses inter-orbital charge fluctuations and amplifies the orbital decoupling that generates the OSQP[7,9-11,14,17]. One approach to identifying such orbital-selective strong correlations experimentally, would be to demonstrate that $Z(\boldsymbol{k})$ is highly distinct between the regions of the electronic bands that are associated with each different orbital $m$.

Because Fe-based materials supporting iron-based superconductivity are excellent candidates to exhibit Hund's metal orbital-selective effects, focus has naturally turned to detecting and understanding such phenomena in these systems. The resulting plethora of theoretical predictions[6,8,11,13,15] include: (i) the electronic structure of iron-based superconductors should be heavily influenced by orbital-selective strong correlations, (ii) this effect is caused primarily by the Hund's decoupling of the interorbital charge fluctuations, (iii) the strength of correlations in each decoupled band $k_j(E)$ grows as it approaches half filling and, (iv) when orbital-selective strong correlations exist in such a state, Cooper pairing itself may become orbital-selective[6,18-21]. Recent photoemission studies of orbital dependent bandwidth renormalization in these materials[22] has been interpreted in this way. However a capability to directly visualize the orbital selectivity of the quasiparticles in the normal state of Fe-based HTS materials, ideally simultaneously with visualization of the electronic structures of the superconducting and nematic phases[3,4], remains an urgent priority.

To address this challenge, we focus on the compound FeSe, which shows clear indications of orbital selectivity[6,15]. The FeSe crystal unit cell has $a$= 2.67 Å, $b$=2.655 Å and $c$= 5.49 Å in the orthorhombic/nematic phase below $T_S \cong 90K$. Specifics of the Fe-plane of the same lattice can be described using the two inequivalent Fe-Fe distances $a_{Fe}$=2.665 Å and $b_{Fe}$=2.655 Å. The Fermi surface (FS) consists of three bands for which an accurate tight-binding model has been developed[18]. This model has excellent simultaneous consistency with angle resolved photoemission[23-25], quantum oscillations[26-28], and Bogoliubov quasiparticle interference[18,19].

Surrounding the Γ=(0,0) point is an ellipsoidal hole-like $\alpha$-band, whose FS $\boldsymbol{k}_\alpha(E = 0)$ has its major axis aligned to the orthorhombic *b*-axis; surrounding the X=(π/a$_{Fe}$,0) point is the electron-like $\varepsilon$-band whose "bowtie" FS $\boldsymbol{k}_\varepsilon(E = 0)$ has its major axis aligned to the orthorhombic *a*-axis; surrounding the Y=(0,π/b$_{Fe}$) point, a $\delta$-band FS should also exist but has proven difficult to detect by spectroscopic techniques. Moreover, it was recently realized that orbital-selective Cooper pairing[18,21] of predominantly the *d$_{yz}$* electrons causes the highly unusual superconducting energy gaps $\Delta_\alpha(\vec{k})$ and $\Delta_\varepsilon(\vec{k})$ of FeSe[18,19], from whose structure the FeSe quasiparticle weights are estimated to be $Z_{xy} \sim 0.1$; $Z_{xz} \sim 0.2$; $Z_{yz} \sim 0.8$ (with the other $Z$ values being irrelevant for energies near E=0 ; Ref. 18). The challenge is to discover if all these exotic phenomena are indeed caused by the existence of orbital-selective strong correlations in a Hund's metal normal state of FeSe.

Imaging of quasiparticle scattering interference[29] is an attractive approach. QPI has become widely used to determine exotic electronic structure of correlated electronic materials[30-35]. This effect occurs when an impurity atom/vacancy scatters quasiparticles which then interfere quantum-mechanically to produce characteristic modulations of the density-of-states $\delta N(\boldsymbol{r}, E)$ surrounding each impurity site; the global effects of this random impurity scattering are usually studied by using $\delta N(\boldsymbol{q}, E)$, the Fourier transform of $\delta N(\boldsymbol{r}, E)$. In a multi-orbital context, this can be predicted using

$$\delta N(\boldsymbol{q}, E) = -1/\pi\, Tr\left(Im \sum_{\boldsymbol{k}} \hat{G}_{\boldsymbol{k}}(E)\, \hat{T}(E)\, \hat{G}_{\boldsymbol{k}+\boldsymbol{q}}(E)\right), \quad (1)$$

where $G_{nm\boldsymbol{k}} = \sqrt{Z_n Z_m} G^0_{nm\boldsymbol{k}}$ with $\hat{G}^0_{\boldsymbol{k}}(E) = \left((E + i\eta)\hat{I} - \hat{H}^{tb}_{\boldsymbol{k}}\right)^{-1}$ is the electron's Green's function in orbital space, $Z_m$ is the quasiparticle weight of orbital $m$, and $\hat{T}(E)$ is a matrix representing all the possible scattering processes between states $|\boldsymbol{k}\rangle$ and $|\boldsymbol{k} + \boldsymbol{q}\rangle$ for an impurity with on-site potential. Atomic scale imaging of these interference patterns $\delta N(\boldsymbol{r}, E)$ is achieved using spatial mapping of differential tunneling conductance $dI/dV(\boldsymbol{r}, E) \equiv g(\boldsymbol{r}, E)$, and has developed into a high-precision technique for measurement of electronic band structure $\boldsymbol{k}_i(E)$ of strongly correlated electron fluids[31-34]. QPI should be of unique utility in searching for both orbital-selective coherence and spectral weight because: (i) the existence of quantum interference itself is a robust test of *k*-space coherence and (ii) the amplitude of QPI signals is sensitive as the squares of quasiparticle weights (Eqn. 1). Our target is thus to achieve orbitally resolved QPI from which the relative $Z_m$ values of the normal state quasiparticles can be estimated.

We pursue this objective in the iron-based superconducting compound FeSe. Fig. 1a is a schematic representation of the orbitally resolved band structure of FeSe at $k_z$=0 (Ref. 18). Surrounding the Γ= (0,0) point, the evolution of $\boldsymbol{k}_\alpha(E)$ is hole-like with the band top near E=+15meV and $d_{yz}$ orbital character (green) maximum along the *x*-axis while $d_{xz}$ orbital character (red) prevails along the *y*-axis. Centered on the X=(π/$a_{Fe}$,0) point, $\boldsymbol{k}_\varepsilon(E)$ exhibits electron-like evolution with two Dirac points near E=-25meV, and $d_{yz}$ orbital character (green) dominant along the *y*-axis while $d_{xy}$ orbital character (blue) prevails along the *x*-axis. A fully coherent δ-band at the Y=(0,π/$b_{Fe}$) point would then have $d_{xz}$ orbital character (red) dominant along the *x*-axis and $d_{xy}$ orbital character (blue) prevailing along the *y*-axis. Fig. 1c,g show the orbitally-resolved constant-energy-contours (CEC) $\boldsymbol{k}_\alpha(E = -10\ meV)$ and $\boldsymbol{k}_\varepsilon(E = +10\ meV)$ of the α- and ε-bands in Fig. 1a. Fig. 1e,i then show the expectations based on Eq. (1) for the intraband QPI intensity patterns $|\delta N_\alpha(\boldsymbol{q}, E = -10\ meV)|$ and $|\delta N_\varepsilon(\boldsymbol{q}, E = +10\ meV)|$ corresponding to these contours, if all |$\boldsymbol{k}$⟩ states are equally and fully coherent. If, by contrast, orbital-selective quasiparticles exist in FeSe, QPI should be very different because the quasiparticle weights $Z_m$ associated with the Fe *d*-orbitals could all be distinct. In that situation, one might expect to see phenomena exemplified schematically by Fig. 1b. Here, for didactic purposes, we have chosen $Z_{xy} < Z_{xz} \ll Z_{yz}$. This means that in $\boldsymbol{k}_\alpha(E)$ the quasiparticle weight of $d_{yz}$ orbital character (green) along the *x*-axis dominates strongly over the quasiparticle weight of $d_{xz}$ orbital character (translucent red) along the *y*-axis (Fig. 1d). Similarly, for $\boldsymbol{k}_\varepsilon(E)$ the quasiparticle weight of $d_{yz}$ orbital character dominates strongly along the *y*-axis compared to the negligible quasiparticle weight of the $d_{xy}$ orbital character (pale blue) along the *x*-axis (Fig. 1h). The δ-band exhibits feeble quasiparticle weight of $d_{xz}$ orbital character along the *x*-axis and negligible $d_{xy}$ quasiparticle weight along the *x*-axis. Under these circumstances, the QPI patterns will obviously be very different because scattering between regions with $Z_m \ll 1$ will produce far weaker intensity modulations. Thus, Fig. 1f,j show the anticipated intraband QPI intensity patterns $|\delta N_\alpha(\boldsymbol{q}, E = -10\ meV)|$ and $|\delta N_\varepsilon(\boldsymbol{q}, E = +10\ meV)|$, when the $|\boldsymbol{k}\rangle$ states have quasiparticle weights $Z_{xy} < Z_{xz} \ll Z_{yz}$. These are obviously quite different than those expected of fully coherent CEC in Fig. 1e,i and for the obvious reason that weak QPI intensity is produced by the quasiparticles of $d_{xz}$ orbital character and virtually none by those of $d_{xy}$ orbital character (SM Section II).

For FeSe, quantitative comparison of the QPI signature $\delta N(\boldsymbol{q}, E)$ expected for fully coherent bands versus strong orbital selectivity of quasiparticles, can then be carried out by using the T-matrix formalism. Here, the fully coherent Greens function $\hat{G}_{\boldsymbol{k}}^0(E)$ representing each $|\boldsymbol{k}\rangle$ state (a 5 by 5 matrix retaining orbital content information) is computed directly from the parameters of the electron band structure (Fig. 1a). These $\hat{G}_{\boldsymbol{k}}^0(E)$ are then used to calculate $\delta N(\boldsymbol{q}, E)$ from Eqn. 1. A scattering matrix $\hat{T}(E) = V_{imp}\hat{I}\left(1 - V_{imp}\sum_{\boldsymbol{k}}\hat{G}_{\boldsymbol{k}}^0(E)\right)^{-1}$ representing $Z_m = 1$ for all $m$ and a δ-function scattering potential at the origin in real space, and only $|\boldsymbol{k}\rangle$ for which $k_z = 0$, are used (SM Section II). Additionally, we numerically calculate the Fourier transform amplitude of the Feenstra transform, $L(\boldsymbol{r}, E) = N(\boldsymbol{r}, E)/\int_0^E N(\boldsymbol{r}, E')dE'$, to compare directly to measured normalized conductance, (dI/dV)/(I/V) (see below and SM Section II). The resulting $|L(\boldsymbol{q}, E)|$ for fully coherent FeSe $|\boldsymbol{k}\rangle$ states are shown in Fig. 2a-d (and in Supplement Movie M1). These $|L(\boldsymbol{q}, E)|$ comprise QPI of α-, ε-, and δ-band for low **q** scattering events. They show all the salient QPI features of fully coherent bands. By contrast, the QPI signatures of an OSQP in FeSe are determined using Eqn. 1 but with $G_{nm}(k, E) = \sqrt{Z_n}\sqrt{Z_m}G_{nm}^0(k, E)$ where $Z_m \in (0.073, 0.94, 0.16, 0.85, 0.36)$ for $m \in (d_{xy}, d_{x^2-y^2}, d_{xz}, d_{yz}, d_{z^2})$ and $V_{imp}$ same as before (SM Section II). (Although these specific values chosen were taken from Ref. 18, the data in this paper as well as the data in Ref. 18 are consistent with the orbitally selective ansatz within a range of Z values that are all consistent with the inequality $Z_{xy} < Z_{xz} \ll Z_{yz}$.) Most relevant are the orbitally resolved quasiparticle weights $Z_{xy} \approx 0.1$; $Z_{xz} \approx 0.2$; $Z_{yz} \approx 0.8$ with the other two orbitals having negligible spectral weight near E=0. The predicted $|L(\boldsymbol{q}, E)|$ for OSQP are shown in Fig. 2i-l (and in Supplement Movie M2). These $L(\boldsymbol{q}, E)$ are now dominated by QPI of both α- and ε-bands as scattering in the δ-band is strongly suppressed due to decoherence of the respective quasiparticles. For the OSQP scenario, the scattering intensity distribution is strikingly $C_2$ symmetric. As expected, the QPI is dominated by quasiparticles with $d_{yz}$ orbital content which are oriented along the $k_x$-axis in the α-band for E<0 while being concentrated along the $k_y$-axis in the ε-band for E>0. This produces the marked rotation of the QPI pattern by 90-degrees just above the chemical potential, a remarkable effect characteristic of FeSe[34] whose origin has until now proven elusive. Clearly the QPI predictions for OSQP (Fig. 2i-l and movie M2) are vividly different than those expected of a fully coherent conventional band structure (Fig. 2a-d and movie M1).

Our experimental search for OSQP phenomena uses spectroscopic imaging scanning tunneling microscopy (SI-STM) to study FeSe. The samples are inserted into the SI-STM instrument and cleaved in cryogenic ultrahigh vacuum at T<20K. To focus on the normal state of FeSe, measurements for the energy range $-8.75 \, meV$ to $+8.75 \, meV$ are acquired at 10.0K > $T_C$, while the rest of the measurements are acquired at 4.2K to reduce thermal smearing. We have checked that the observed QPI phenomena do not differ between 4.2K and 10.0K (see SM section VIII). Differential tunneling conductance $g(\boldsymbol{r}, E) \equiv dI/dV(\boldsymbol{r}, E = eV)$ measurements are carried out with atomic resolution and register, as a function of both location $\boldsymbol{r}$ and electron energy $E$. Because of the tiny areas of FeSe bands in $\boldsymbol{k}$-space (Fig. 1a), intraband QPI wavevectors are limited $|\boldsymbol{q}(E)| < 0.25(\frac{2\pi}{a_{Fe}})$, so that high-precision $g(\boldsymbol{r}, E)$ imaging in very large fields of view (typically 50X50 nm²) is required. The Fourier transform of $g(\boldsymbol{r}, E)$, $g(\boldsymbol{q}, E)$, can then be used to reveal wavevectors and intensities of dispersive modulations due to QPI. However, to avoid artifacts (SM Section III) images of $L(r, E = eV) \equiv \left(\frac{g(r,E)}{I(r,E)}\right)V$ are more typically used, and these faithfully portray relative intensity at different directions in $q$-space[34]. Thus, Figs 2e-h show the measured $|L(q,E)|$, the Fourier transform amplitude of $L(r,E)$, from FeSe samples where the only scattering defects in the FOV are at Fe sites (topograph of measurement FOV is shown in SI Section IV and $|L(q,E)|$ is provided as Supplement Movie). All such QPI data rotate by 90-degrees when measurements of $|L(q,E)|$ are made in the orthogonal orthorhombic domain (SM Section V). Comparison of the measured QPI in Figs 2e-h to predicted $|L(q,E)|$ for fully coherent bands (Figs 2a-d) and for OSQP (Figs 2i-l) reveals that the latter are in far better agreement. The intensity pattern and energy dispersion of the $q$-vectors of maximum scattering intensity in measured $|L(q,E)|$ closely follow those shown in Figs 2i-l, including the strong unidirectionality, and the sudden rotation of dispersion direction as E=0 is crossed. This provides a direct signature of OSQP in the metallic state of FeSe.

To visualize the impact of orbital selectivity on the complete band structure more globally, one can compare the energy dispersions continuously by comparing computed $|L(q_x,E)|$ and $|L(q_y,E)|$ to measured $|L(q_x,E)|$ and $|L(q_y,E)|$ respectively. For this purpose, Fig. 3a shows the theoretical dispersion of QPI maxima for α-, ε-, and δ-band along both $q_x$ and $q_y$ resolved by orbital content using the same color code as elsewhere. Fig. 3b shows the energy dependence of the predicted intensity of intraband scattering interference, along the same two trajectories as in Fig. 3a for fully coherent quasiparticle weights in all three orbitals $Z_{xy} = Z_{yz} = Z_{xz} = 1$. Fig. 3c shows the measured intensity of intraband scattering interference along $q_x$ and $q_y$. The correspondence of these data to predictions in Fig. 3b is quite poor. However, in Fig. 3d we show the predicted intensity of intraband scattering interference if FeSe exhibits orbital selective QPI. The same two $E$-$q$ planes as in Fig. 3b,c are shown, but now the OSQP quasiparticle weights are $Z_{xy} \approx 0.1$; $Z_{xz} \approx 0.2$; $Z_{yz} \approx 0.8$. The correspondence between experimental $|L(\boldsymbol{q}, E)|$ (Fig. 3c) and the QPI signature of OSQP (Fig. 3d) is good and is discernibly superior to that with Fig. 3b.

If the quasiparticle weights indeed obey the relation $Z_{yz} \gg Z_{xz} > Z_{xy}$, it begs the question of whether weak QPI can be observed on the δ-band for its sections dominated by $d_{xz}$ orbital content. Such phenomena should be clearest at states E>10meV (because the QPI from the α-band has disappeared here) and should appear along $q_x$ due to scattering interference between $d_{xz}$ dominated quasiparticles connected by a double-headed arrow shown in Fig. 4a. As seen in Fig. 4b, the expected scattering of states on the $δ$ pocket is significantly suppressed. The remaining panels of Fig. 4 demonstrate that there is indeed a dispersive signal along $q_x$ at somewhat higher $q$ than the significantly stronger scattering interference along $q_y$ from the $d_{yz}$ sections of the ε-band. Detailed analysis and comparison of these two electron-like dispersive signals to simulation allow the conclusion that even the $d_{xz}$ orbital content quasiparticles with very low $Z_{xz}$ are detectable, as expected, on the δ-band (SM Section VI).

Finally, to visualize approximately how the $Z_j(\boldsymbol{k})$ evolve with $\boldsymbol{k}$-space angle around the Fermi surfaces of the α- and ε-bands, we measure the magnitude of $L(\boldsymbol{q}, E)$ on the $\boldsymbol{q}$-space trajectory through the QPI data for both bands. Fig. 5 a,b, show the measured angular dependence of QPI intensity for intraband scattering within the α– and ε-bands. The assignment of the scattering intensity to the electron and hole bands can be made by observing the dispersion of the intensity as a function of energy. In both cases, we focus on the trajectory of ***q=2k*** intraband scattering as indicated by the white crosses at which a local maximum in QPI amplitude is detected; the data are shown in full detail versus energy in SM Section VII. The $L(\boldsymbol{q}, E)$ amplitude is determined by taking line cuts through the measured $|L(\boldsymbol{q}, E)|$ maps (Fig. 5a,b) for a sequence of angles at a specified energy. Each line cut was fit to a sum of a linear background and a Gaussian peak to determine QPI signal amplitude (SM Section VII). The analysis was carried out for a sequence of energies (-25 meV to -15 meV for the α-band and +15 meV to +25 meV for the ε-band in 1.25 meV steps), and then the mean of these amplitudes (black dots in Fig. 5c,d) was taken over the relevant energy range ; the error bars represent the standard deviation of an amplitude as energy is varied. Fig. 5c shows the measured $L(\boldsymbol{q}, E)$ intensity of α-band intraband QPI versus the $\boldsymbol{q}$-space angles from Fig. 5a integrated over energy range -25meV≤E≤-15meV where this band is clear and distinct. Comparison to the theoretically predicted $L(\boldsymbol{q}, E)$ intensity (blue dot-dash curve) versus $\boldsymbol{k}$-space angle for orbitally selective QPI with $Z_{xy} \approx 0.1$; $Z_{xz} \approx 0.2$; $Z_{yz} \approx 0.8$ (Section VII), reveals good agreement. Similarly, comparison of the measured ε-band $L(\boldsymbol{q}, E)$ versus the $\boldsymbol{q}$-space angles from Fig. 5b to the predicted $L(\boldsymbol{q}, E)$ intensity (blue dot-dash curve Fig. 5d) for orbitally selective quasiparticles having the same $Z_{xy}:Z_{xz}: Z_{yz}$ ratios (SM Section VII), yields $Z_{xy}$~0. Therefore, the measured $L(\boldsymbol{q}, E)$ amplitude of QPI data (Fig. 5) are strongly consistent with orbital selectivity in the Hund's' metal quasiparticles of FeSe for which $Z_{xy} < Z_{xz} \ll Z_{yz}$.

The measured $Z_m$ phenomena in Fig. 2-5 reveal the strength of orbitally selective strong correlations in the normal metal sate of FeSe. The data indicate that this metal has delocalized $|\boldsymbol{k}\rangle$ states of $d_{yz}$ character with good coherence because $Z_{yz}$~1, $|\boldsymbol{k}\rangle$ states of $d_{xz}$ character that are significantly less coherent, and $|\boldsymbol{k}\rangle$ states of $d_{xy}$ character with lowest relative coherence. Comparison of measured $|L(\boldsymbol{q}, E)|$ to the theoretical $|L(\boldsymbol{q}, E)|$ predictions for different ratios $Z_{xy}:Z_{xz}:Z_{yz}$ (Fig. 2), along with evaluation of the $\boldsymbol{k}$-angle dependence of the QPI intensity for both bands (Fig. 5) (SM Section VII) all indicate that $Z_{xy} < Z_{xz} \ll Z_{yz}$. Moreover, we find the ratio of quasiparticle weights $Z_{xy}:Z_{xz}:Z_{yz}$ producing good agreement between theoretical $|L(\boldsymbol{q}, E)|$ and the QPI data $|L(\boldsymbol{q}, E)|$ (Figs. 2, 3, 4) to be indistinguishable from those deduced independently from the energy gap structure caused by orbital-selective Cooper pairing[18,19]. This provides strong support for the concept of orbital-selective quasiparticle identification and $Z$ quantification using QPI. Of most significance is that these orbital-selective QPI data provide direct demonstration that the normal state from which the HTS emerges in FeSe is dominated by orbitally selective strong correlations. If true in general for the iron-based HTS materials, this would be of fundamental significance because strong electronic correlations would then play a central role in both copper-based and iron-based high temperature superconductivity.

**Acknowledgements**: We are grateful to S.D. Edkins, A. Georges, M.H. Hamidian, J.E. Hoffman, G. Kotliar, E.-A. Kim, D.-H. Lee, L. de Medici, P. Phillips, and J.-H. She for helpful discussions and communications. J.C.S.D. and P.C.C. gratefully acknowledge support from the Moore Foundation's EPiQS (Emergent Phenomena in Quantum Physics) Initiative through grants GBMF4544 and GBMF4411, respectively. P.J.H. acknowledges support DOE Grant No. DE-FG02-05ER46236. A.Kr. and BMA acknowledge support from a Lundbeckfond Fellowship (Grant No. A9318). Material synthesis and detailed characterization at Ames National Laboratory was supported by the U.S. Department of Energy, Office of Basic Energy Science, Division of Materials Sciences and Engineering - Ames Laboratory is operated for the U.S. Department of Energy by Iowa State University under Contract No. DE-AC02-07CH11358; Experimental studies were carried out by the Center for Emergent Superconductivity, an Energy Frontier Research Center, headquartered at Brookhaven National Laboratory were funded by the U.S. Department of Energy under DE-2009-BNL-PM015.

**Author Contributions:** A.K., Y.X.C. and P.O.S. developed and carried out the experiments; A.E.B. and P.C. synthesized and characterized the samples; A.K., P.O.S., and A.Kr., developed and carried out analysis; A.Kr., B.M.A. and P.J.H. provided theoretical guidance; B.M.A , P.J.H. and J.C.S.D. supervised the project and wrote the paper with key contributions from A.K., Y.X.C., P.O.S., A.Kr, and P.J.H. The manuscript reflects the contributions and ideas of all authors.

**Additional Information:** The data described in the paper are archived by the Davis Research Group at Cornell University and can be made available by contacting the corresponding author.

**Competing financial interests:** The authors declare no competing financial interests.

Figure 1

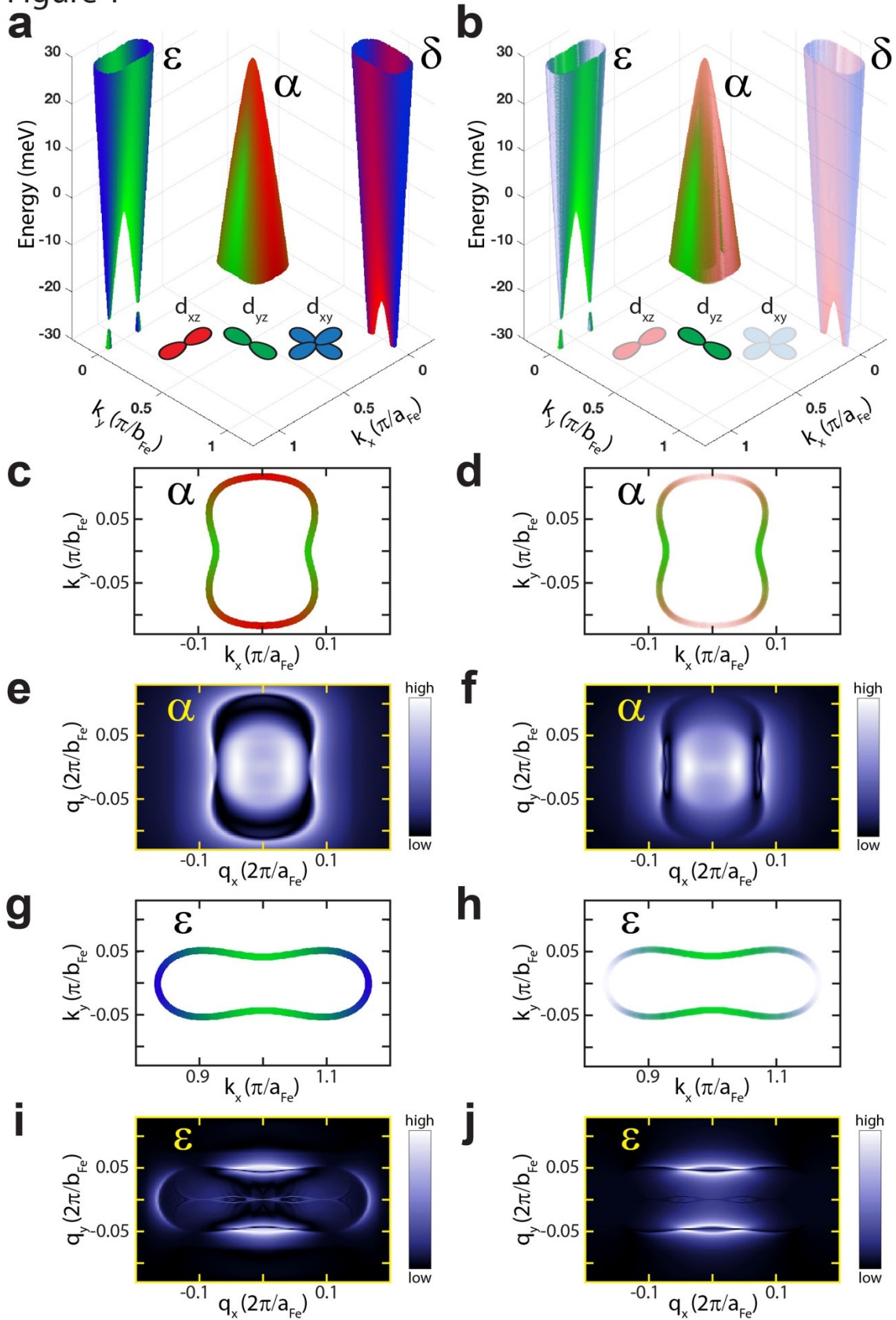

**Figure 1    Orbitally resolved quasiparticle scattering interference in FeSe**

a. Schematic representation of orbitally resolved band structure of FeSe at $k_z=0$. For each fully coherent quasiparticle state $|k\rangle$ in these bands green represents $d_{yz}$ orbital content, red represents $d_{xz}$ orbital content and blue represents $d_{xy}$ orbital content. The two Dirac points on the band surrounding the X-point ($\pi/a_{Fe}$,0) occur near $E = -25\ meV$ while the top of the hole-like band surrounding the Γ-point (0,0) is close to $E = +15\ meV$

b. Schematic representation of the same orbitally resolved band structure of FeSe at $k_z=0$ but now indicating the effects of different quasiparticle weight Z. Here green represents the virtually fully coherent $d_{yz}$ orbital content, translucent red represents the reduced Z value of $d_{xz}$ orbital content and pale blue represents $d_{xy}$ orbital content where Z tends towards zero.

c. Orbital content of constant-energy-contours (CEC) at the Γ-point (0,0) at -10 meV using same color code as **a**.

d. Orbital content of CEC at the Γ-point (0,0) at -10 meV using same color code as **b**.

e. Anticipated $|\delta N_\alpha(q, E = -10\ meV)|$ QPI signature of intraband scattering interference within α-band surrounding the Γ-point, for quasiparticle weights $Z_{xy} = Z_{yz} = Z_{xz} = 1$. The $|\delta N(q,E)|$ ($\equiv |1/\pi\ Tr\ Im \sum_k \hat{G}_k(E)\ \hat{T}(E)\ \hat{G}_{k+q}(E)|$) images in panels E,F, I and J are calculated using T matrix with weak impurity potential using the band structure model displayed in panels **a** and **b**. In the calculations, the **k** sum was restricted to the appropriate region of the Brillouin zone to separately capture intraband scattering interference pattern for different pockets.

f. Anticipated $|\delta N_\alpha(q, E = -10\ meV)|$ QPI signature for α-band with orbital-selective quasiparticle weights $Z_{xy} \approx 0.1$; $Z_{xz} \approx 0.2$; $Z_{yz} \approx 0.8$.

g. Orbital content of CEC at the X-point ($\pi/a_{Fe}$,0) at +10 meV using same color code as **a**.

h. Orbital content of CEC at the X-point ($\pi/a_{Fe}$,0) at +10 meV using same color code as **b**.

i. Anticipated $|\delta N_\varepsilon(\boldsymbol{q}, E = +10\ meV)|$ QPI signature of intraband scattering interference within ε-band for quasiparticle weights $Z_{xy} = Z_{yz} = Z_{xz} = 1$.

j. Anticipated $|\delta N_\varepsilon(\boldsymbol{q}, E = +10\ meV)|$ QPI signature for ε-band with orbital-selective quasiparticle weights $Z_{xy} \approx 0.1$; $Z_{xz} \approx 0.2$; $Z_{yz} \approx 0.8$.

Figure 2

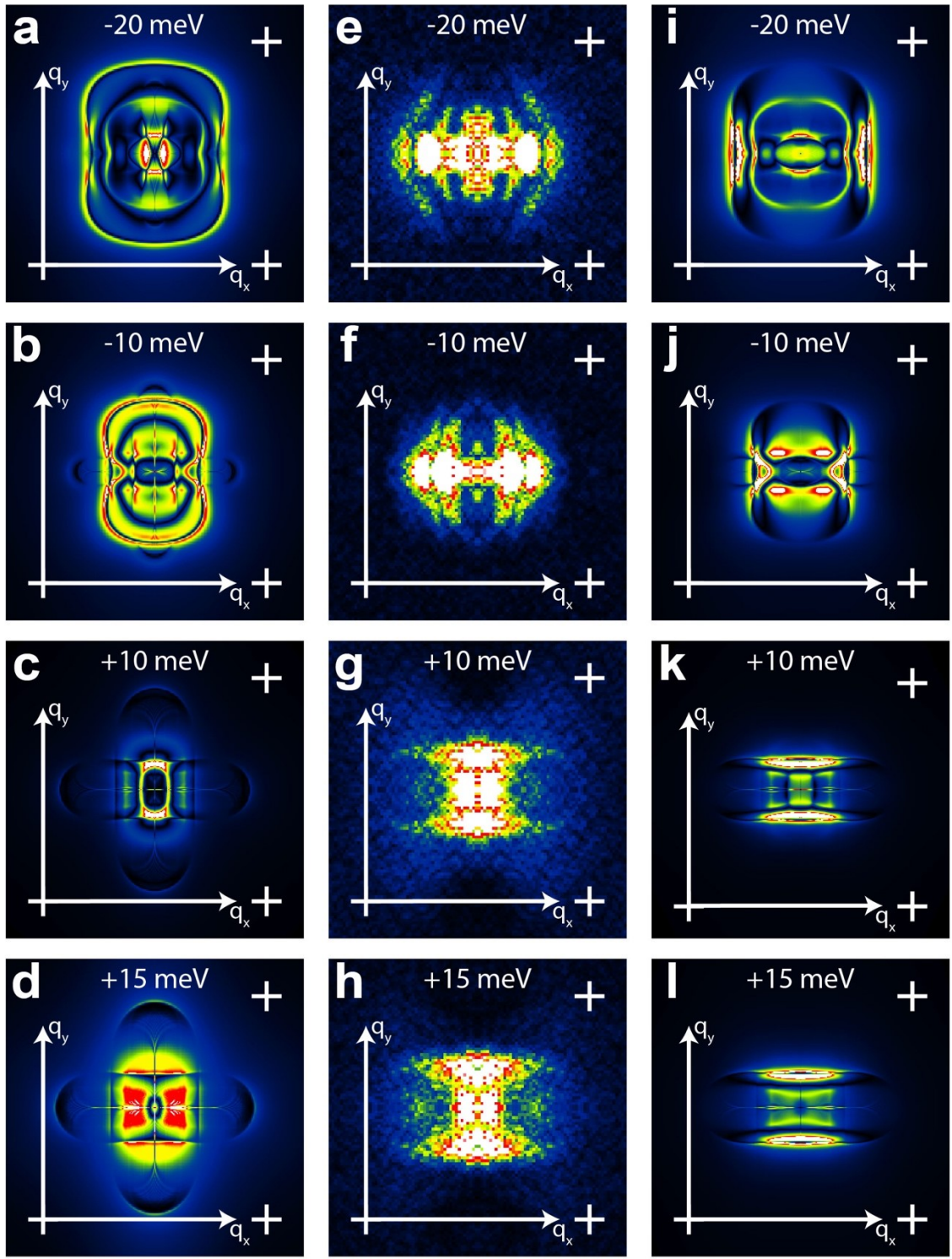

**Figure 2 Visualizing orbital-selective quasiparticle interference**

**a-d** Predicted energy-resolved $|L(\boldsymbol{q},E)|$ QPI signature of intraband scattering interference in a fully coherent state for quasiparticle weights $Z_{xy} = Z_{yz} = Z_{xz} = 1$. The white crosses correspond to $\frac{3}{16}\left(\frac{2\pi}{a_{Fe}}, \frac{2\pi}{b_{Fe}}\right)$ points in the momentum space. $|L(\boldsymbol{q},E)|$ is the amplitude of the Fourier transform of the normalized conductance ($\equiv (\frac{dI}{dV})/(\frac{I}{V})$) at wavevector $\boldsymbol{q}$ and energy $E$.

**e-h** Measured $|L(\boldsymbol{q},E)|$ of FeSe at the same energies as shown in **a-d** and **i-l**. For all these energies, the measurements agree much better with the orbital-selective quasiparticle (OSQP) scenario (**a-d**) than with the fully coherent QPI predictions (**i-l**).

**i-l** Predicted energy-resolved $|L(\boldsymbol{q},E)|$ QPI signature of intraband scattering interference in a OSQP with quasiparticle weights $Z_{xy} \approx 0.1$; $Z_{xz} \approx 0.2$; $Z_{yz} \approx 0.8$.

Figure 3

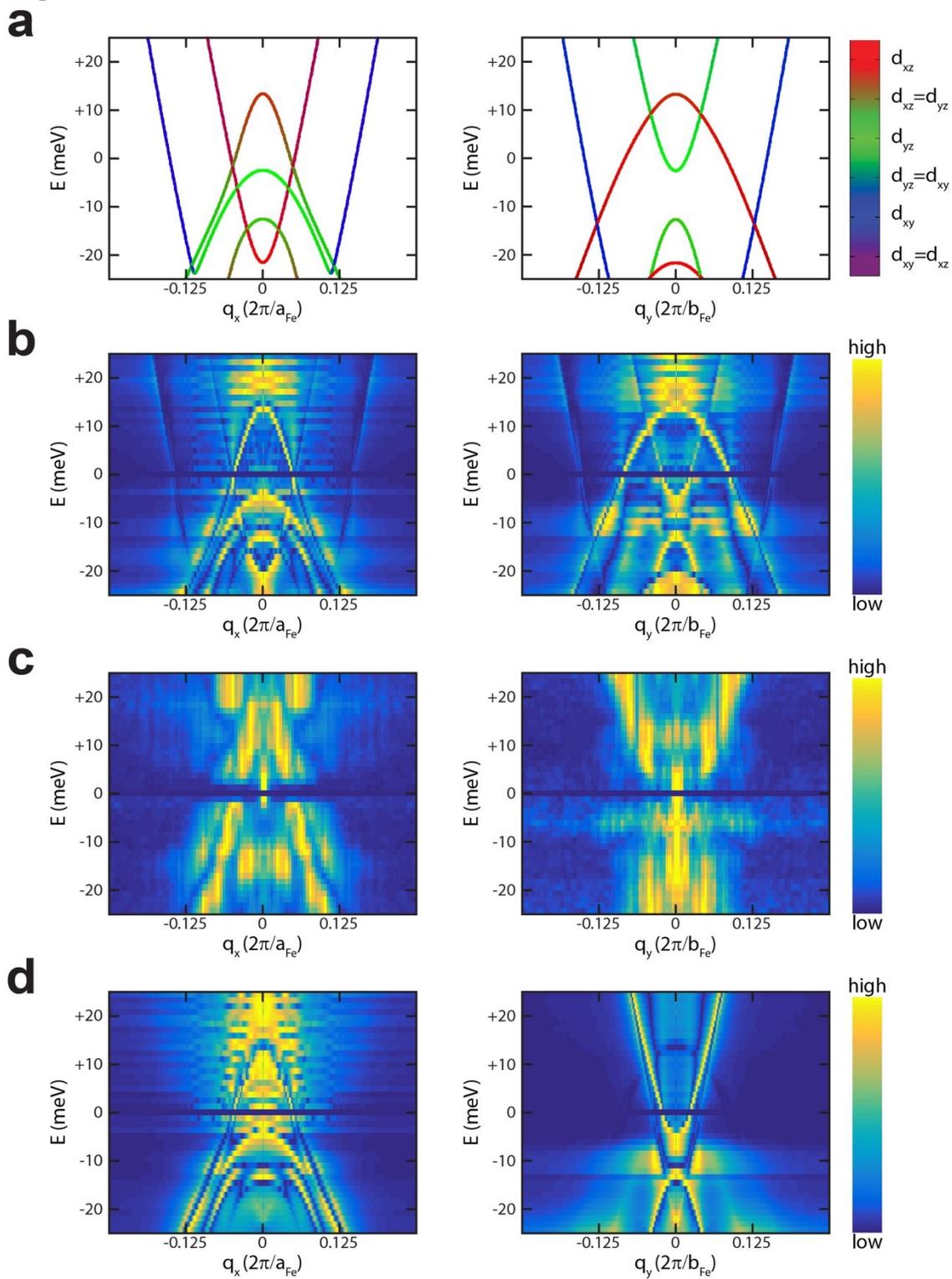

**Figure 3 Energy dependence of orbital-selective quasiparticle interference**

a. Momentum space representation of intraband quasiparticle interference maxima resolved by orbital content. Two E-q (energy-wavevector) planes are shown, parallel to $q_x$ and to $q_y$. Color code shows the orbital content.

b. Predicted intensity of intraband scattering interference for a fully coherent state. Two E-q planes are shown, parallel to $q_x$ and to $q_y$. The quasiparticle weights are $Z_{xy} = Z_{yz} = Z_{xz} = 1$.

c. Measured intensity of intraband scattering interference in FeSe. Same two E-q planes as in **b** are shown. Correspondence of these data to predictions in **b** is poor, whereas their correspondence to the orbital-selective QPI prediction in **d** is much better.

d. Predicted intensity of intraband scattering interference for orbital-selective quasiparticles (OSQP) in FeSe. Same two E-q planes as in **b** are shown. The OSQP quasiparticle weights here are $Z_{xy} \approx 0.1$; $Z_{xz} \approx 0.2$; $Z_{yz} \approx 0.8$.

Images in panels **b,c**, and **d** are generated from $q_x$ and $q_y$ line cuts of the corresponding calculated and measured $|L(\boldsymbol{q}, E)|$. These cuts are normalized to unity for each energy to enhance visibility of band dispersions.

Figure 4

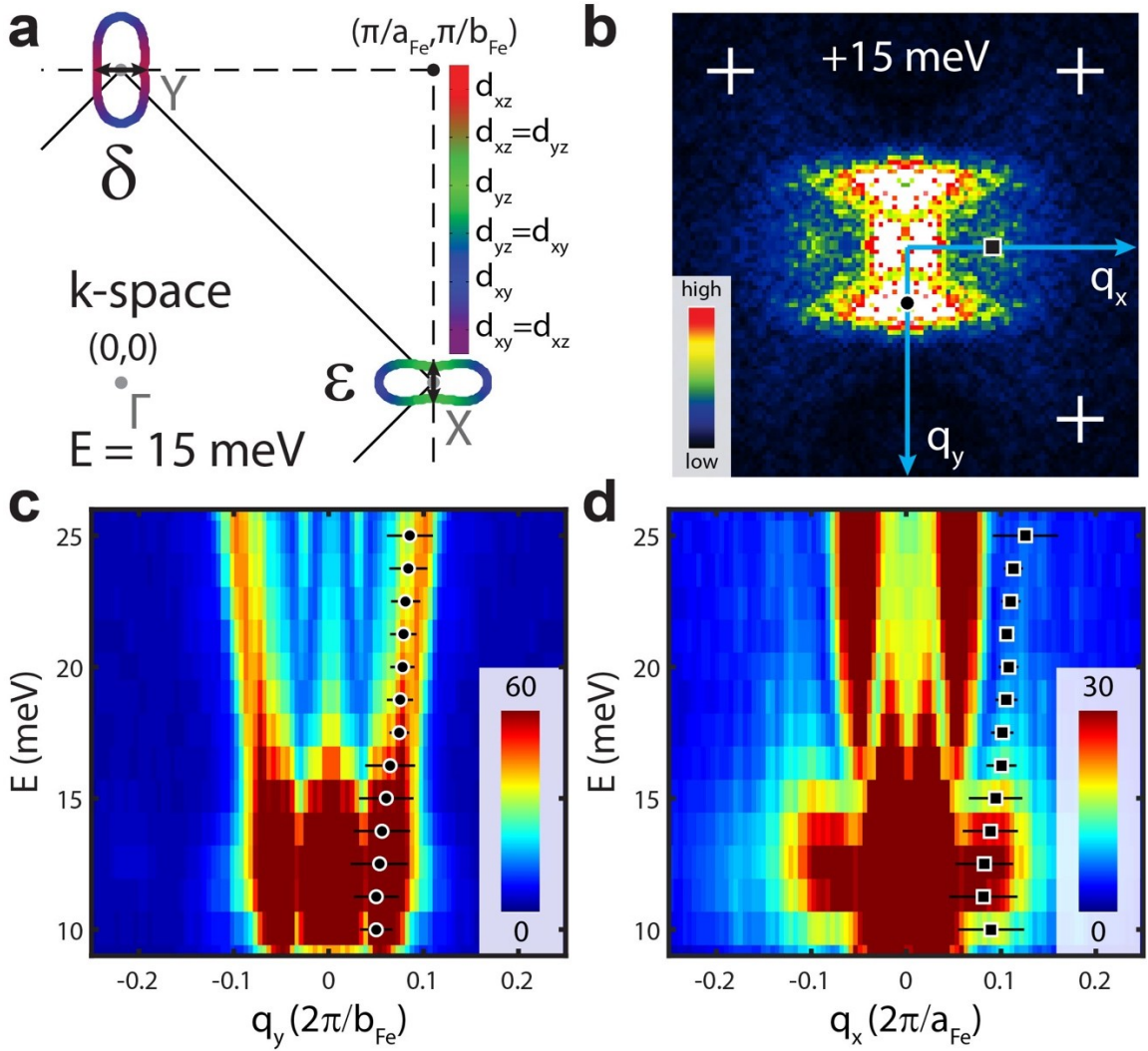

**Figure 4 Detecting orbital-selective quasiparticle interference from both ε- and δ-bands above $E_f$**

a. Quasiparticle constant energy contours at $E = +15\ meV$ showing the ε- and δ-bands. The color code indicates whether the quasiparticles are dominated by $d_{yz}$ (green), $d_{xz}$ (red) or $d_{xy}$ (blue) orbital character. Double-headed arrows show scattering vectors along $q_x$ and $q_y$.

b. Measured $|L(\boldsymbol{q}, E = +15\ meV)|$ image. The directions of $q_x$ and $q_y$ line cuts are shown as blue lines. The signals from $\varepsilon$ and $\delta$ bands are marked by a black circle and a black square, respectively. Signal locations were determined from fits to the line cuts (SI section VI). The white crosses correspond to $\frac{3}{16}\left(\frac{2\pi}{a_{Fe}}, \frac{2\pi}{b_{Fe}}\right)$ points in q space. $|L(\boldsymbol{q}, E)|$ is the amplitude of the Fourier transform of the normalized conductance ($\equiv (\frac{dI}{dV})/(\frac{I}{V})$) at wavevector $\boldsymbol{q}$ and energy $E$.

c. E-$q_y$ line cut through the sequence of measured $|L(\boldsymbol{q}, E)|$ images. The line cuts were fit to Gaussian peaks, and the locations of the peaks and the corresponding widths are shown as black circles with black lines.

d. E-$q_x$ line cut through the sequence of measured $|L(\boldsymbol{q}, E)|$ images. The line cuts were fit to Gaussian peaks, and the locations of the peaks and the corresponding widths are shown as black squares with black lines. Note that the maximum intensity in **d** is 50% of the maximum intensity in **c** with respect to the color bars.

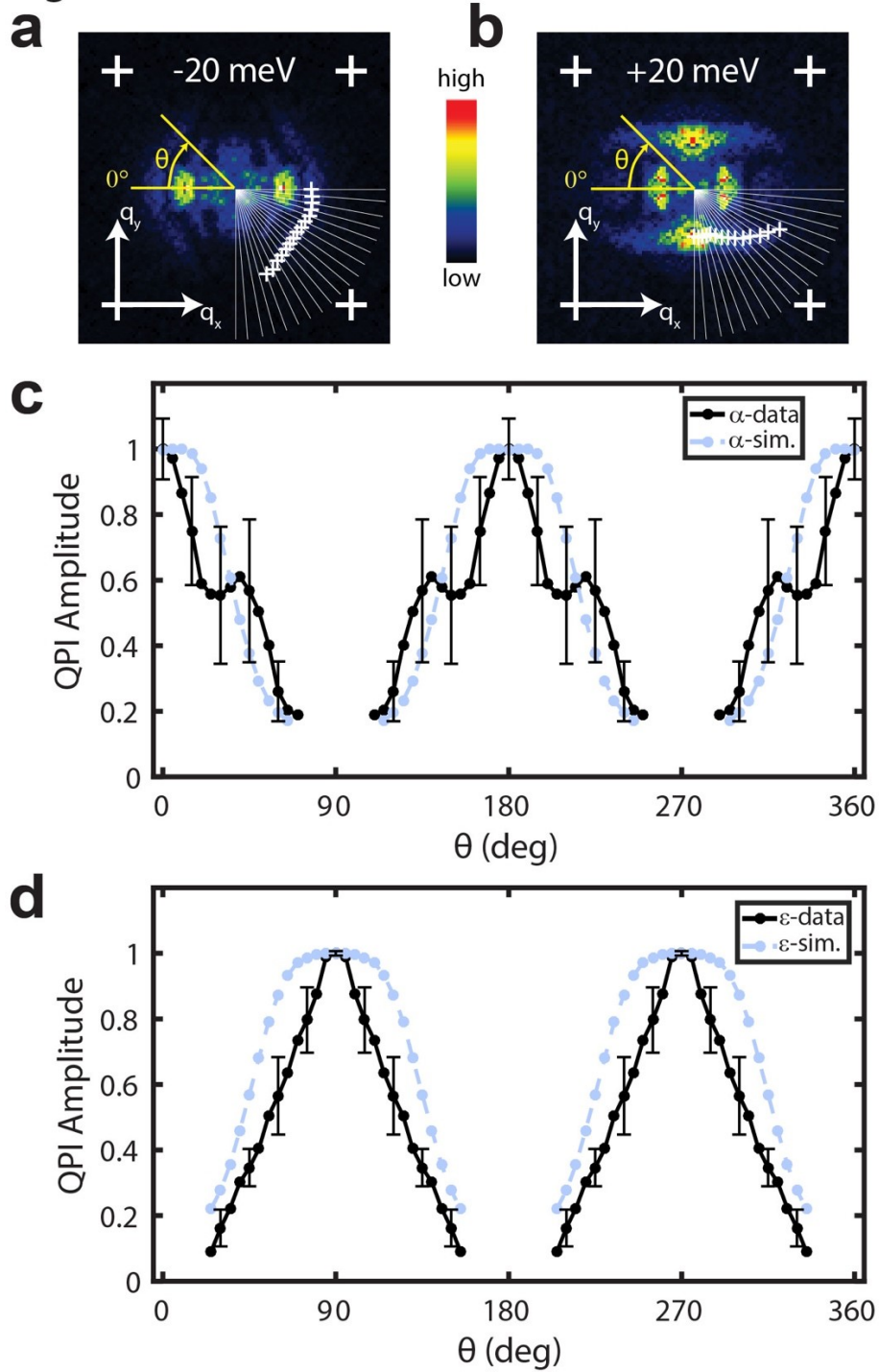

**Fig. 5 Momentum-angle dependence of orbital-selective quasiparticle weight $Z_m$**

a. Measured $|L(\boldsymbol{q}, E = -20\ meV)|$ image from the $\alpha$-band showing the trajectory of the angularly resolved line cuts in **c**. Small white crosses mark the extracted

peak location of QPI intensity, see SI section VII. The large white crosses correspond to $\frac{3}{16}\left(\frac{2\pi}{a_{Fe}}, \frac{2\pi}{b_{Fe}}\right)$ points in momentum space. $|L(q,E)|$ is the amplitude of the Fourier transform of the normalized conductance ($\equiv \left(\frac{dI}{dV}\right)/\left(\frac{I}{V}\right)$) at wavevector $q$ and energy $E$.

b. Measured $|L(q, E = +20\ meV)|$ image from the $\varepsilon$-band showing the trajectory of angularly resolved line cuts in **d**. Small white crosses mark the extracted peak location of QPI intensity, see SI section VII.

c. Measured mean $L(q,E)$ amplitude versus angle for the $\alpha$ band. The amplitudes are extracted from measured $|L(q,E)|$ images on trajectory shown by crosses in **a**, and averaged over -25 to -15 meV energy range with the error bar showing the standard deviation for the sequence of amplitudes at different energies. Blue symbols show the predicted values of $|L(q, E = -20\ meV)|$ in the orbital-selective quasiparticle scenario with quasiparticle weights $Z_{xy} \approx 0.1$; $Z_{xz} \approx 0.2$; $Z_{yz} \approx 0.8$.

d. Measured mean $L(q,E)$ amplitude versus angle for the $\varepsilon$ band. The amplitudes are extracted from measured $|L(q,E)|$ images on trajectory shown by crosses in **b**, and averaged over 15 to 25 meV energy range with the error bar showing the standard deviation for the sequence of amplitudes at different energies. Blue symbols show the predicted values of $|L(q, E = +20\ meV)|$ in the orbital-selective quasiparticle scenario with quasiparticle weights $Z_{xy} \approx 0.1$; $Z_{xz} \approx 0.2$; $Z_{yz} \approx 0.8$.

# Supplementary Materials for
## Imaging Orbital-selective Quasiparticles
## in the Hund's Metal State of FeSe

A. Kostin, P.O. Sprau, A. Kreisel, Yi Xue Chong, A. E. Böhmer, P.C. Canfield, P.J. Hirschfeld, B.M. Andersen and J.C. Séamus Davis

correspondence to: jcseamusdavis@gmail.com

## Supplementary Text

This paper deals with orbital-selective quasiparticle spectral weight in the Hund's metal state of iron based superconductors. The importance of multi-orbital effects for the physics of strongly correlated electron systems was originally noted in DMFT calculations that raised a possibility of orbital-selective Mott transitions (OSMT)[1-3]. The pioneering theoretical studies focused on OSMT[3-10] while more recent research has focused on predictions of the Hund's metal state[11]. Since then, the importance of the Hund's metal state for iron based superconductors has been pointed out in detail, for example in Refs 5-12 of the main text.

### Band structure model summary and comment about the choice of axis

For all the analysis and theoretical QPI simulations included in this paper, we use a band structure parameterization of FeSe introduced in (*13,14*). Specifically, the relevant Hamiltonian is $\hat{H}_k^{tb} = \hat{H}_0 + \hat{H}_{OO} + \hat{H}_{SOC}$, where $\hat{H}_0$ (in real space notation) is given by

$$\hat{H}_0 = \sum_{r,r',a,b} t_{r-r'}^{ab} c_{a,r}^\dagger c_{b,r'} \qquad (S1)$$

where $a, b$ are orbital labels and $r, r'$ are lattice sites. For the orbital order term, we use the momentum space representation,

$$\hat{H}_{OO} = \Delta_b(T) \sum_{k} \bigl(\cos(k_x) - \cos(k_y)\bigr)\bigl(n_{xz}(\mathbf{k}) + n_{yz}(\mathbf{k})\bigr) + \Delta_s(T) \sum_{k} \bigl(n_{xz}(\mathbf{k}) - n_{yz}(\mathbf{k})\bigr) \qquad (S2)$$

Finally, the spin orbit coupling is given by

$$\hat{H}_{SOC} = \lambda \mathbf{L} \cdot \mathbf{S} \qquad (S3)$$

For best fit to quantum oscillation, angle resolved photoemission and quasiparticle interference data, the parameter values are $\Delta_s = 9.6$ meV, $\Delta_b = -8.9$ meV, and the SOC constant is fixed to $\lambda = 20$ meV.

We would like to draw the reader's attention to the choice of lattice axes and coordinates used in this paper. In Fig. S1, we display the schematic of the FeSe crystal structure in the nematic state where the orthorhombic distortion has been exaggerated for clarity.

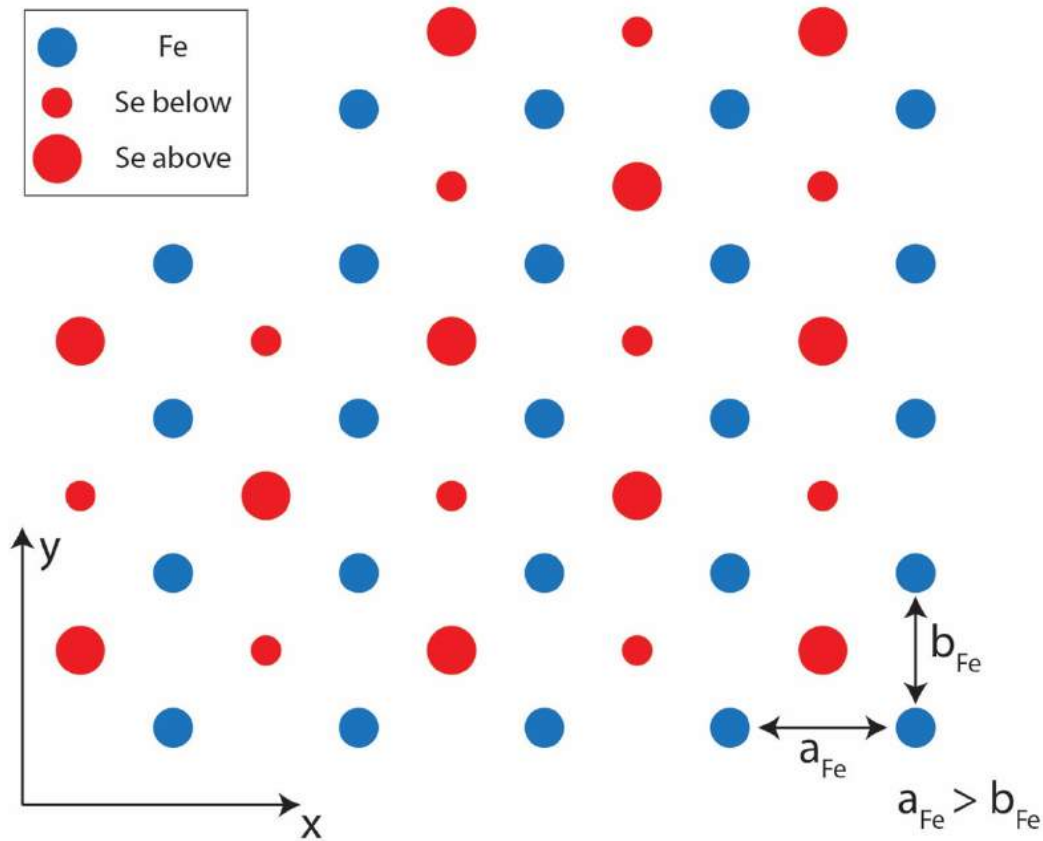

**Figure S1| FeSe crystal structure**

We choose the convention $a_{Fe} > b_{Fe}$ for the Fe-Fe nearest neighbor distances with the x-axis pointing along $a_{Fe}$. This convention is consistent with previous superconductivity studies on this material (*13,15*). Recently, another STM study on FeSe$_{1-x}$S$_x$ chose the opposite convention (*16*). The measurements in that study are consistent with this work with the trivial replacement a ↔ b and x ↔ y (with respect to the orbital labels in the context of the conclusions of this paper this would lead to the interchange $d_{xz}$ ↔ $d_{yz}$).

1. **T matrix calculation details (parameters, quasiparticle weights)**

The theoretical predictions of quasiparticle interference patterns presented in Fig. 2 of the main paper were performed within the T-matrix approach. We model a constant on-site scatterer at position $\boldsymbol{r}^*$ by adding the impurity term to the Hamiltonian, $\widehat{H}_{imp} = V_{imp} \sum_a c_{a,\boldsymbol{r}^*}^\dagger c_{a,\boldsymbol{r}^*}$. From the unperturbed Green's function, $\widehat{G}_{\boldsymbol{k}}^0(\omega) = \left([\omega + i\eta]\widehat{I} - \widehat{H}_{\boldsymbol{k}}^{tb}\right)^{-1}$, we obtain the Green's function in the presence of an impurity, $\widehat{G}_{\boldsymbol{k},\boldsymbol{k}'}(\omega) = \widehat{G}_{\boldsymbol{k}-\boldsymbol{k}'}^0(\omega) + \widehat{G}_{\boldsymbol{k}}^0(\omega)\widehat{T}(\omega)\widehat{G}_{\boldsymbol{k}'}^0(\omega)$. Here we use the tight binding parameterization specified above in Section 1 for $\widehat{H}_{\boldsymbol{k}}^0$ and a value of 1.25 meV for the energy smearing parameter, $\eta$. The value for the smearing was chosen to be consistent with the modulation voltage of the lock-in used for the dI/dV measurements. The T-matrix is given by $T(\omega) = \frac{V_{imp}\widehat{I}}{\widehat{I} - V_{imp}\sum_{\boldsymbol{k}}\widehat{G}_{\boldsymbol{k}}^0(\omega)}$. The matrix representing the impurity potential was chosen proportional to identity with $V_{imp} = -100\ meV$. Then within this formalism, the change in the local density of states due to impurity scattering is $\delta N(\boldsymbol{q}, \omega) = -\frac{1}{\pi} Tr\{Im \sum_{\boldsymbol{k}} \widehat{G}_{\boldsymbol{k}}^0(\omega) \widehat{T}(\omega) \widehat{G}_{\boldsymbol{k}+\boldsymbol{q}}^0(\omega)\}$.

To implement orbitally selective correlations, we introduce the orbitally dependent quasiparticle weights $Z_m$ where $m \in (d_{xy}, d_{x^2-y^2}, d_{xz}, d_{yz}, d_{z^2})$ into the Green's function $G_{nm\boldsymbol{k}} = \sqrt{Z_n Z_m} G_{nm\boldsymbol{k}}^0$.

Due to the setup effect, present in the experimental measurement of the local density of states, we use the Feenstra function to make comparisons between measurements and T matrix calculations. See Section 3 below for details. The predicted Feenstra function can be readily computed numerically within the T matrix approach in the following sequence of steps.

1) Compute the Inverse Fourier Transform of $\delta N(\boldsymbol{q}, \omega)$ to go to real space.

$$\delta N(\boldsymbol{r}, \omega) = \sum_{\boldsymbol{q}} e^{i\boldsymbol{q}\cdot\boldsymbol{r}}\, \delta N(\boldsymbol{q}, \omega) \qquad (S4)$$

2) Add the uniform component of the density of states to the perturbed one.

$$N(\boldsymbol{r}, \omega) = N^0(\omega) + \delta N(\boldsymbol{r}, \omega) = -\frac{1}{\pi} Tr\left\{Im \sum_{\boldsymbol{k}} G_{\boldsymbol{k}}^0(\omega)\right\} + \delta N(\boldsymbol{r}, \omega) \qquad (S5)$$

3) Compute the Feenstra function in real space.

$$L(\boldsymbol{r}, \omega) = \frac{N(\boldsymbol{r}, \omega)}{\sum_{\omega'=0}^{\omega'=\omega} N(\boldsymbol{r}, \omega')} \qquad (S6)$$

4) Calculate the Feenstra function in $q$-space by Fourier transform.

$$L(\boldsymbol{q}, \omega) = \sum_{\boldsymbol{r}} e^{-i\boldsymbol{q}\cdot\boldsymbol{r}} L(\boldsymbol{r}, \omega) \tag{S7}$$

## 2. The setup effect in quasiparticle interference measurements & Feenstra parameter

The STM tunneling current is thought to be well approximated by the following equation (*17*):

$$I(V, \boldsymbol{r}) = \frac{4\pi e}{\hbar} e^{-s\sqrt{\frac{8m\varphi}{\hbar^2}}} n_t(0) \int_{-eV}^{0} n_s(\epsilon, \boldsymbol{r}) d\epsilon \tag{S8}$$

Here $s$ is the tip sample separation, $\varphi$ is the tunnel barrier height (some admixture of the work functions of the tip and the sample), and $\boldsymbol{r} = (x, y)$. Hence the sample density of states $n_s(\epsilon, \boldsymbol{r})$ is defined explicitly as a function of the lateral $(x, y)$ coordinates.

Now using the standard lock-in technique, experimentalists can easily measure $dI/dV$ as well. Using the equation above, we evaluate this observable below to conclude that it is proportional to the sample density of states.

$$\frac{dI}{d(-eV)}(V, \boldsymbol{r}) = -\frac{4\pi e}{\hbar} e^{-s\sqrt{\frac{8m\varphi}{\hbar^2}}} n_t(0) n_s(-eV, \boldsymbol{r}) \tag{S9}$$

We establish the junction at a specific location $(x, y)$ at a particular tunnel barrier width $s$ by specifying a certain current $I_0$ at certain bias $V_0$ in feedback.

$$I_0 = \frac{4\pi e}{\hbar} e^{-s\sqrt{\frac{8m\varphi}{\hbar^2}}} n_t(0) \int_{-eV_0}^{0} n_s(\epsilon, \boldsymbol{r}) d\epsilon \tag{S10}$$

Since the STM measurements are performed at constant $(I_0, V_0)$, then it follows that $s$ is a function of the lateral coordinates, $s = s(\boldsymbol{r})$.

Using this equation, we can rewrite the top two formulae as follows.

$$I(V, \boldsymbol{r}) = I_0 \frac{\int_0^{-eV} n_s(\epsilon, \boldsymbol{r}) d\epsilon}{\int_0^{-eV_0} n_s(\epsilon, \boldsymbol{r}) d\epsilon} \tag{S11}$$

$$g(V, \boldsymbol{r}) = \frac{dI}{dV}(V) = I_0 \frac{-e n_s(-eV, \boldsymbol{r})}{\int_0^{-eV_0} n_s(\epsilon, \boldsymbol{r}) d\epsilon} \tag{S12}$$

Note that the quantity defined as $g(V, \boldsymbol{r})$ is what is usually reported in a lot of STM QPI studies. The denominator $\int_0^{-eV_0} n_s(\epsilon, \boldsymbol{r}) d\epsilon$ in the expressions above depends on x and y since the local density of states $n_s(\epsilon, \boldsymbol{r})$ is a function of x and y. This is known as the setup effect, and it is a major problem since we only want to look at spatial modulations of the density of states at a particular energy, $n_s(-eV, \boldsymbol{r})$. Instead, the experimentally accessible quantity $g(V)$ will have spatially modulating signals from a range of energies because of the integral in the denominator. Note that the exact $g(V, \boldsymbol{r})$ is a function of $V_0$, an arbitrary parameter.

Alternatively, we can define the Feenstra function (*18*) by using both available measurements, $I(V)$ and $\frac{dI}{dV}(V)$.

$$L(V, \boldsymbol{r}) = \frac{\frac{dI}{dV}}{I/V} = V \frac{-en_s(-eV)}{\int_0^{-eV} n_s(\epsilon) d\epsilon} \quad (S13)$$

Here, we have very similar problems that existed in $g(V, \boldsymbol{r})$. Neither $L(V, \boldsymbol{r})$ or $g(V, \boldsymbol{r})$ report modulations of $\rho_s$ at a single energy corresponding to the bias V. However, $L(V, \boldsymbol{r})$ is not a function of $V_0$. Experimentally, it is possible to acquire several $dI/dV$ maps at different $V_0$ in the same field of view, and they will potentially all look different. However, $L(V, \boldsymbol{r})$ will be the same in principle. Reporting SI-STM data in the format $L(V, \boldsymbol{r})$ and $L(V, \boldsymbol{q})$ allows direct comparison of results measured on different instruments and across all different research groups, independent of the completely random setup conditions used for each g(*r*,E) map at any lab. This universality for data inter-comparison is actually a very important motivation to use the format $L(V, \boldsymbol{r})$ and $L(V, \boldsymbol{q})$. Moreover, the QPI measurable quantity $L(E, \boldsymbol{q}) = FT\{\frac{n_s(E,r)}{\int_0^E n_s(E',r)dE'}\}$ can in principle be determined.

3. **Topographic image of the experimental FOV**

Figure S2 presents a high resolution constant current topograph of the field of view used for the QPI study. The main type of defect is attributed to Fe vacancies and produces very strong Friedel oscillations. A second type of defect that could be caused by a vacancy in the Se-lattice is comparatively very rare, and no Friedel response is observed.

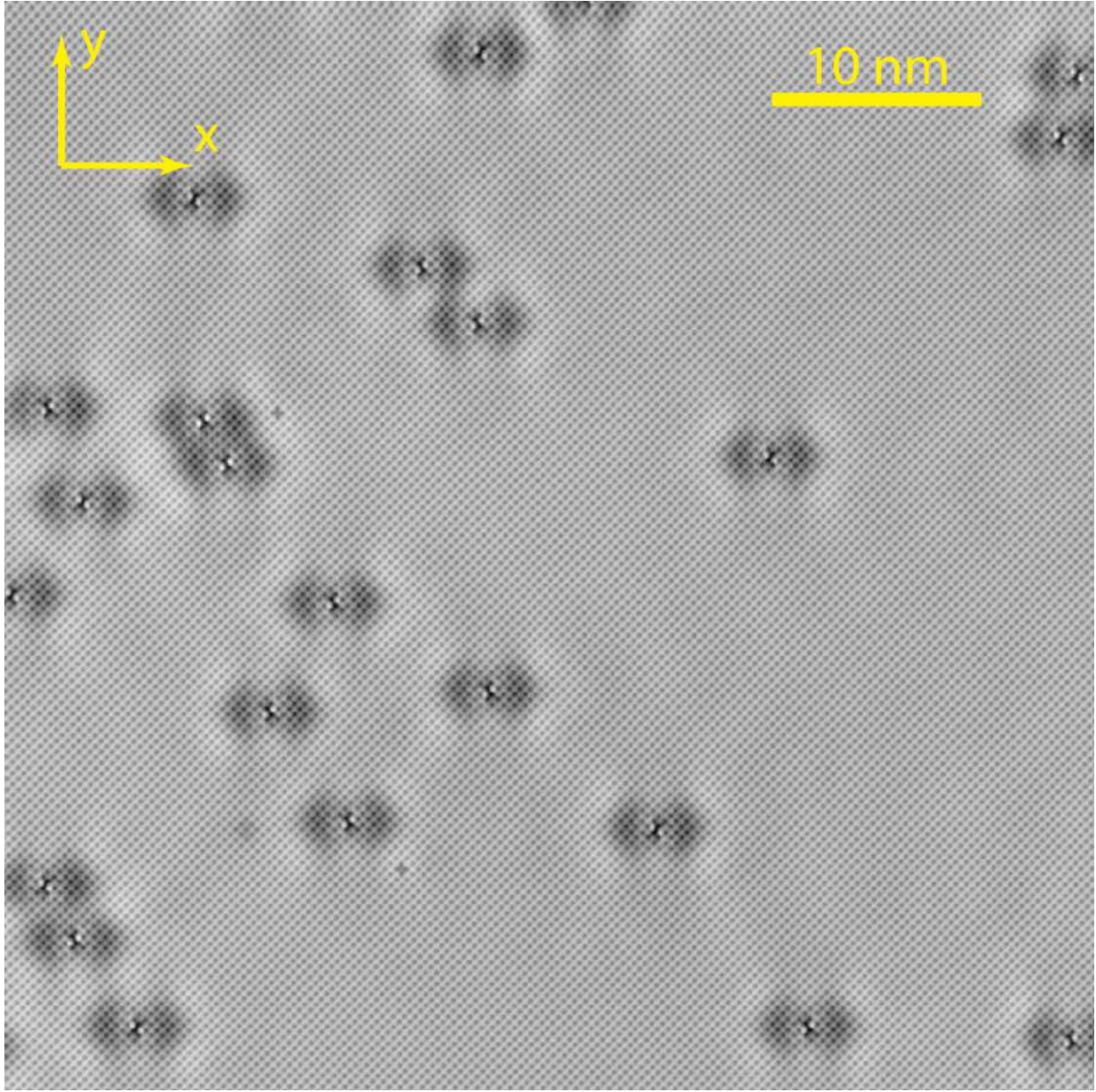

**Figure S2| Topography of the FOV for single domain QPI studies.** The FOV size is 52 nm and the resolution is 1024 pixels. Topography is acquired at -20 mV and 10 pA setup condition.

### 4. QPI rotation between domains

In Fig. S3 below, we show the quasiparticle interference patterns in the proximity of a twin boundary. The quasiparticle interference signal rotates across the orthorhombic twin boundary as expected.

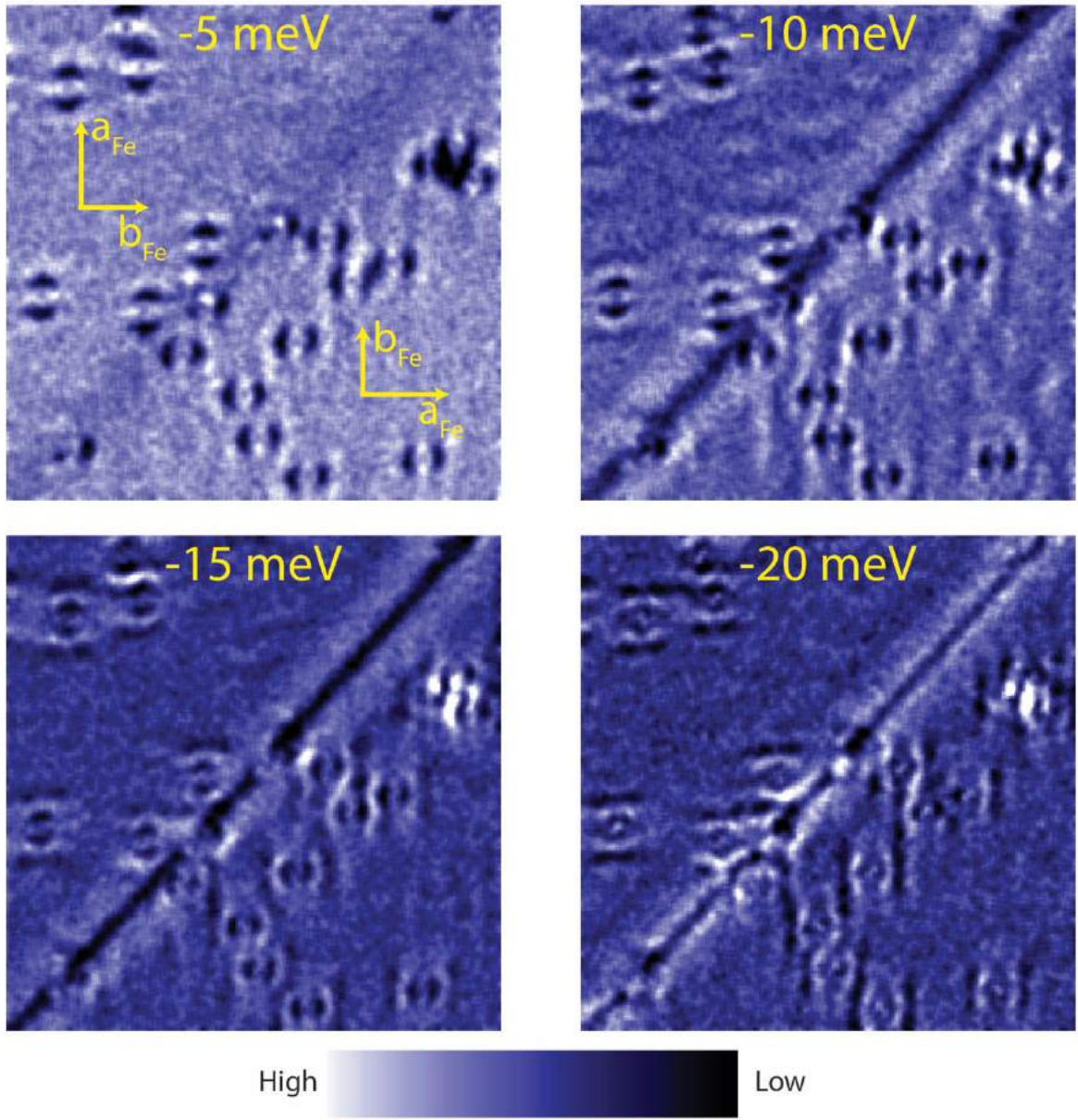

**Figure S3| QPI patterns in the proximity to twin domain boundary.** A sequence of measured $L(\mathbf{r}, E)$ images is displayed with the twin boundary clearly visible. The scattering wave vectors rotate across the twin boundary, and below the chemical potential they are along the $a_{Fe}$ direction. Map resolution is 256 pixels, and the FOV size is 50 nm.

## 5. QPI dispersions and amplitudes for $\varepsilon$ and $\delta$ bands

In the next two sections, we describe and discuss in detail the analysis of highly anisotropic QPI, and how the experimental observations can be explained within the framework of orbital-selective decoherence. In the energy range of this study, the three relevant Fe-orbitals are $d_{xz}$, $d_{yz}$, and $d_{xy}$.

If quasiparticle weights of these orbitals markedly differ intraband scattering intensity of the electron and hole pockets will be strongly dependent on their orbital content. This allows us to approach the problem from two different directions. First, we focus on a comparison of the electron pockets $\delta$ and $\varepsilon$ which are dominated by $d_{xz}$- + $d_{xy}$- and $d_{yz}$- + $d_{xy}$-content, respectively (see Fig 1 in the main text). In section 7, we analyze the detailed angular dependence of QPI intensity for intraband scattering of the hole pocket $\alpha$ and the electron pocket $\varepsilon$, and how it relates to the orbital content as a function of angle.

Let us examine low-**q** QPI in the energy range between 10 and 25 meV above the chemical potential. In this energy range, the low-**q** QPI is dominated by the intraband scattering within the $\delta$- and $\varepsilon$-pockets (see Fig. S4). For the case of a fully coherent metal with equal quasiparticle weights for all three orbitals we expect the two strong signals dispersing along both $q_x$ and $q_y$. The situation is quite different in the orbital-selective scenario with large $d_{yz}$, almost completely suppressed $d_{xy}$, and strongly suppressed $d_{xz}$ quasiparticle weight. A very strong signal disperses along $q_y$, and a weaker signal disperses along $q_x$. The strong signal is caused by scattering between parts of the $\varepsilon$-pocket of predominantly $d_{yz}$ orbital content, and the weak signal corresponds to scattering between parts of the $\delta$-pocket of predominantly $d_{xz}$ orbital content. For both pockets scattering between parts of predominantly $d_{xy}$ orbital content is very strongly suppressed. The observed QPI pattern in experiment agrees very well with the orbital-selective simulation.

Figure S5 presents the remaining experimental |L(**q**, E)| layers. All data has been symmetrized taking advantage of the mirror symmetry axes of the orthorhombic crystal unit cell. In Fig. S6, we show cuts along $q_x$ and $q_y$ through the fully coherent and orbital-selective simulations. Both figures also contain the experimentally extracted position of dispersing signals as black dots with white edges. There is very good agreement between the extracted q-vectors and the expected dispersion based on the tight-binding model.

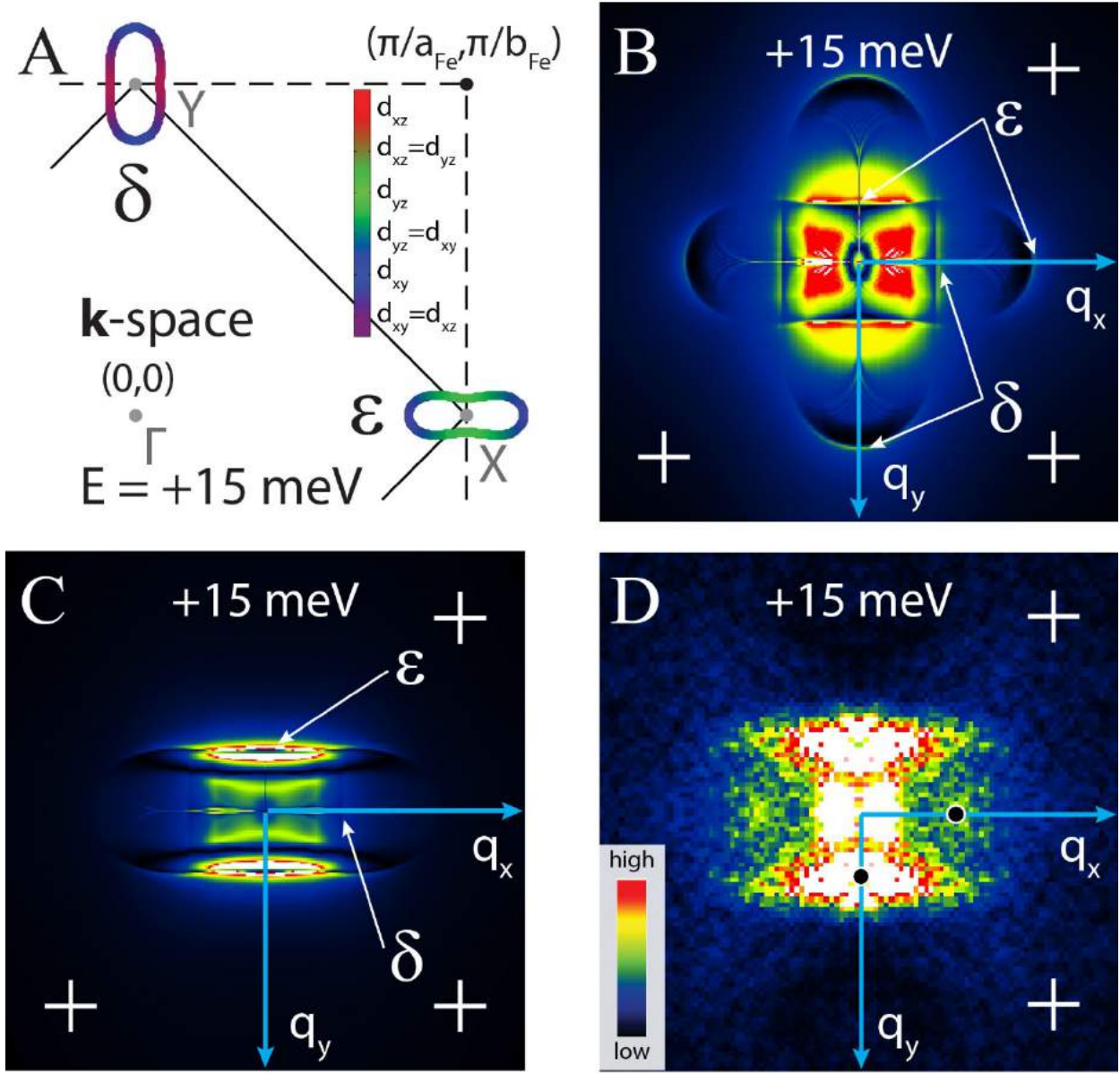

**Figure S4| Experimental and simulated intraband QPI for electron pockets above the chemical potential. A,** Orbitally resolved constant-energy-contour at +15 meV for the tight-binding model depicting the electron-like pockets $\varepsilon$ and $\delta$. The solid line marks the 2-Fe Brillouin zone, and the dashed line represents the 1-Fe Brillouin zone. **B,** Simulated $|L(\mathbf{q}, 15\text{ meV})|$ using equal quasiparticle weights for all orbitals. **C,** Simulated $|L(\mathbf{q}, 15\text{ meV})|$ using orbital-selective quasiparticle weights ($Z_{xy} \sim 0.1$; $Z_{xz} \sim 0.2$; $Z_{yz} \sim 0.8$). **D,** Measured, symmetrized $|L(\mathbf{q}, 15\text{ meV})|$; the black dots with white circles mark the extracted position of the dispersive signal in the line-cuts presented in Fig. S7.

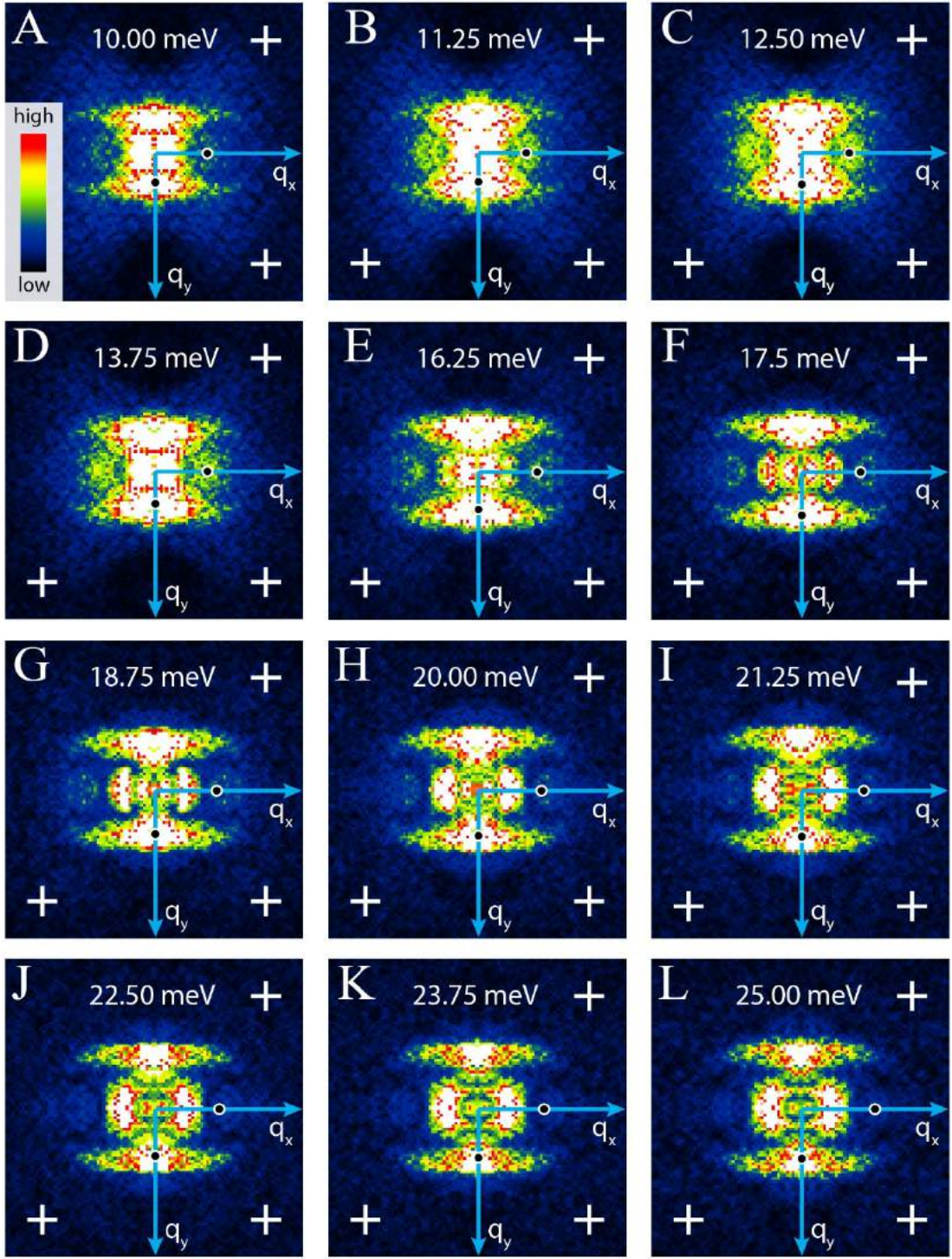

**Figure S5| Anisotropic QPI between 10 and 25 meV. A – L,** Measured, symmetrized |L(**q**, E)| as a function of energy. Blue arrows mark the directions of line-cuts shown in Fig. S7. Black dots with white circles mark the extracted peak location from line-cuts shown in Fig. S7.

We extract the $q$-vectors from line-cuts along $q_x$ and $q_y$ by fitting Gaussian curves with a constant and linear background to the data: $c_0 + c_1(q - q_0) + A \exp\left[-\frac{(q-q_P)^2}{2\sigma^2}\right]$. The line-cuts are depicted in Fig. S7. In order to improve fit quality, the line-cuts are first smoothed by averaging 3 adjacent data points in the initial line-cut. The error bars for the $q$-vectors are given by the full-width-half-maximum parameter of the Gaussian. In addition to the position of the peak, we can analyze its amplitude A. In the orbital-selective scenario it is predicted that intraband scattering of the $\delta$-pocket is suppressed compared to the $\varepsilon$-pocket. This is what we find in experiment, see Fig. S8B.

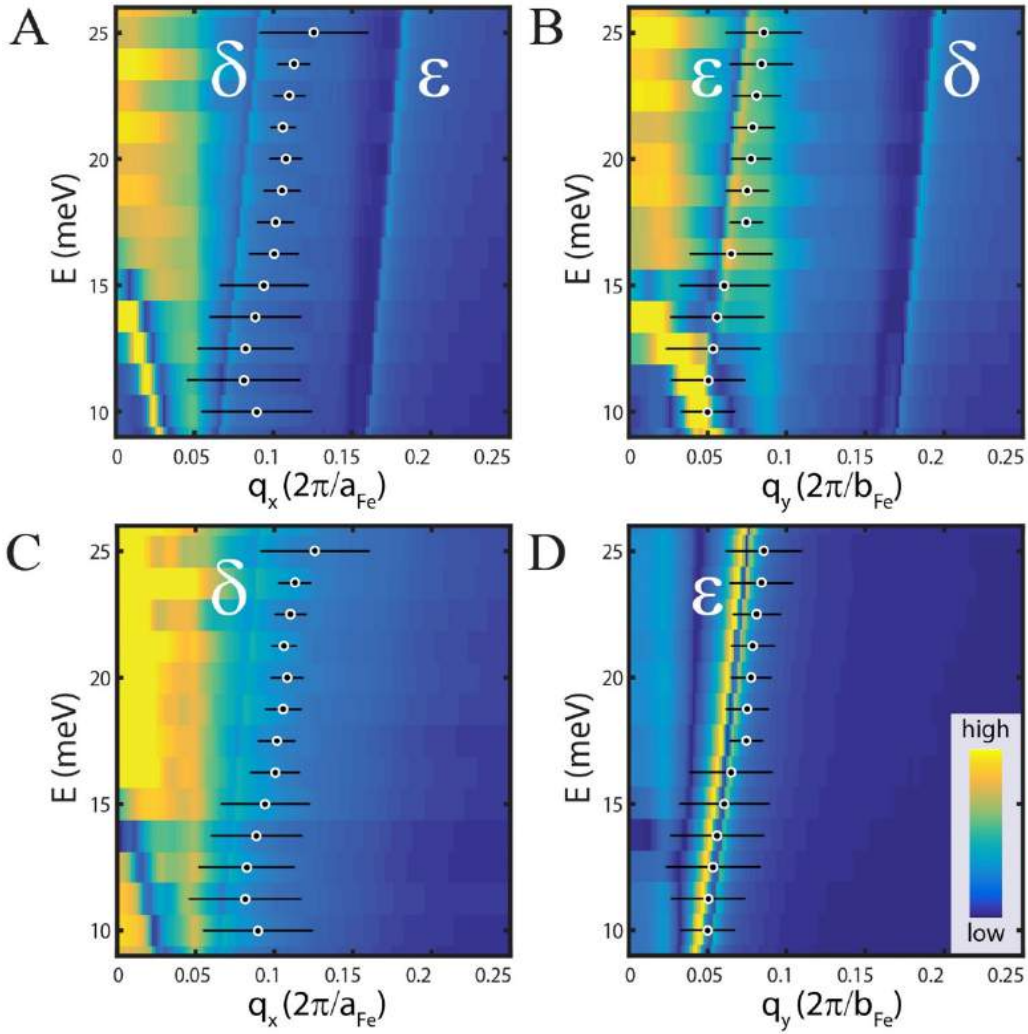

**Figure S6| Comparison of experimental and simulated QPI dispersion for electron pockets. A, B,** Comparison of extracted QPI dispersion to line-cuts through simulation with equal quasiparticle weights. **C, D,** Comparison of extracted QPI dispersion to line-cuts through orbital-selective simulation. The black dots with white circles mark the extracted position of the dispersive signal in the line-cuts presented in Fig. S7.

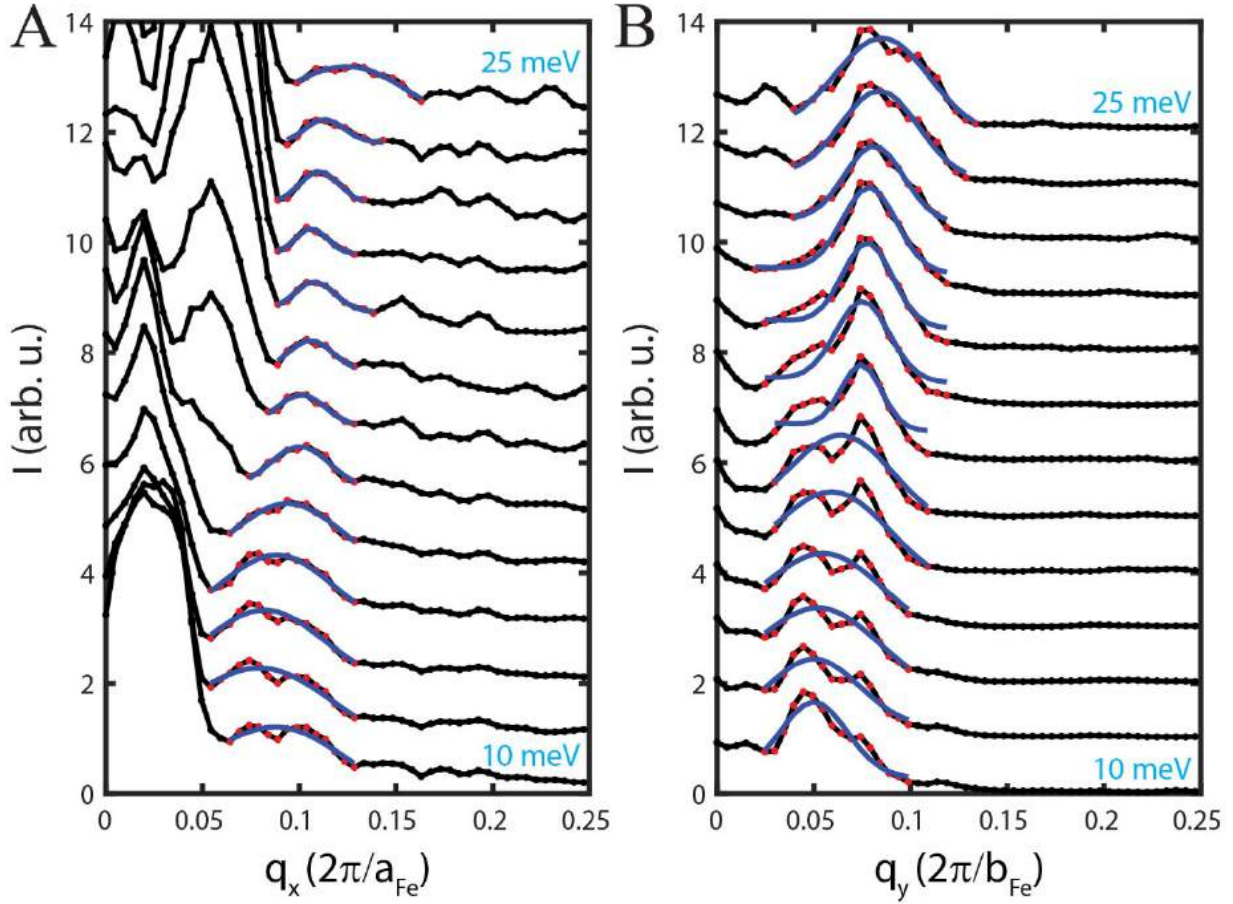

**Figure S7| QPI line-cuts along $q_x$ and $q_y$. A,** Line-cuts along $q_x$ between 10 meV and 25 meV in 1.25 meV steps. **B,** Line-cuts along $q_y$ between 10 meV and 25 meV in 1.25 meV steps. In both A and B, line-cuts have been normalized and shifted in intensity to enhance the visibility of the dispersing peak. Gaussian fitted to the peak is shown in blue.

Figure S8 summarizes the results of our QPI analysis: i) We do not detect any signal related to scattering between regions of dominant $d_{xy}$ content; ii) Scattering between constant-energy-contour parts with predominantly $d_{yz}$ orbital content ($\varepsilon$-pocket) produces significantly stronger QPI intensity than parts where $d_{xz}$ dominates ($\delta$-pocket).

All these findings are consistent with moderate (pronounced) orbital-selective decoherence of the $d_{xz}$- ($d_{xy}$-) orbital compared to the $d_{yz}$-orbital: $Z_{yz} \gg Z_{xz} > Z_{xy}$. The strong decoherence of the $d_{xy}$-orbital was expected from earlier experiments and theoretical studies. The difference between $d_{yz}$ and $d_{xz}$ is probably related to the nematic phase in FeSe which creates a strong anisotropy in the electronic degrees of freedom.

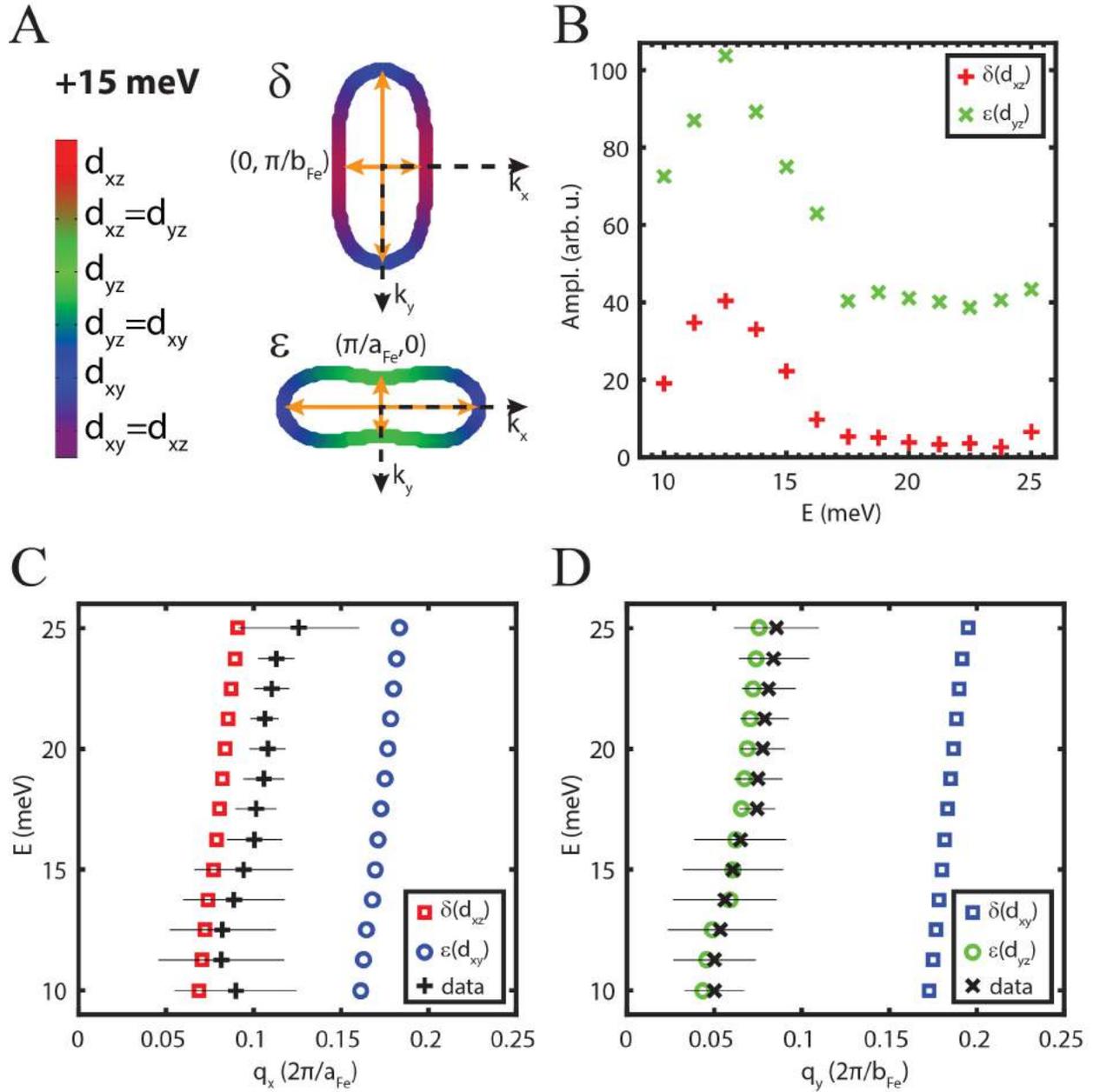

**Figure S8| Summary of line-cut analysis of intra-pocket scattering for electron-like bands above chemical potential. A,** Schematic picture of orbitally resolved intra-pocket scattering space for the two electron pockets at +15 meV. Orange, double-headed arrows symbolize intra-pocket scattering along $q_x$ and $q_y$. **B,** Energy dependence of amplitudes extracted for intra-band scattering of the $\delta$- and $\varepsilon$-pocket. **C, D,** Comparison of extracted QPI dispersion along $q_x$ and $q_y$ from line-cuts in Fig. S7 to extracted QPI dispersion from the simulation for intraband scattering associated with different dominant 3d orbital content.

## 6. QPI amplitude extractions for α-band and ε-band

In addition to comparing the QPI due to intraband scattering of the $\delta$- and $\varepsilon$-pocket along $q_x$ and $q_y$ we explore the complete angular dependence of the QPI intensity for the $\alpha$- and $\varepsilon$-pocket. Figure S9 demonstrates how the intensity decreases as a function of angle in orthogonal fashion for the $\alpha$- and $\varepsilon$-pocket. We studied the angular dependence of the hole pocket $\alpha$ for nine energies (-25 meV to -15 meV in 1.25 meV steps). Likewise, we studied the electron pocket $\varepsilon$ for nine energies (15 meV to 25 meV in 1.25 meV steps). We choose the energy range so that the QPI signals are easily separable.

We show an example of the analysis in Figs. S10, S11. As in section 6, the line-cuts are first smoothed by averaging 3 adjacent data points in the initial line-cut, and the peaks are fitted by a Gaussian with constant and linear background: $c_0 + c_1(q - q_0) + A \exp\left[-\frac{(q-q_P)^2}{2\sigma^2}\right]$. In both cases, the amplitudes clearly decrease as a function of angle until their signal becomes too weak to be detected. We analyze the simulated QPI in a similar fashion. The only difference is that the amplitude is determined by taking the intensity value at the peak position. Since the simulated QPI generates very skewed peaks, a simple Gaussian fit is not possible. Angular dependence of the QPI intensity is analyzed both for the case of orbital-selective (Figs. S12, S13) and equal quasiparticle weights (Figs. S14, S15).

The results are summarized in Fig. S16. Here the experimental values are obtained by taking the average of the nine energies for the $\alpha$- and $\varepsilon$-pocket, respectively. The errorbars represent the computed standard deviation for the nine energies. For the simulation, we only analyze $\pm 20$ meV, as the simulation does not contain any type of noise. It is clear, that the fully coherent simulation with equal quasiparticle weights is inconsistent with the observations in experiment. While experiment shows a strong suppression where $d_{yz}$-orbital content is diminished this is not the case for the coherent simulation.

The additional set of maxima observed in the simulation is due to the geometry of the pockets in the tight-binding model. As a consequence, nesting leads to increased scattering for certain angles which is independent of orbital-selective decoherence. As such it is observed both for the fully coherent and orbital-selective simulation. The dominant effect is nevertheless the influence of the orbital-selective quasiparticle weights.

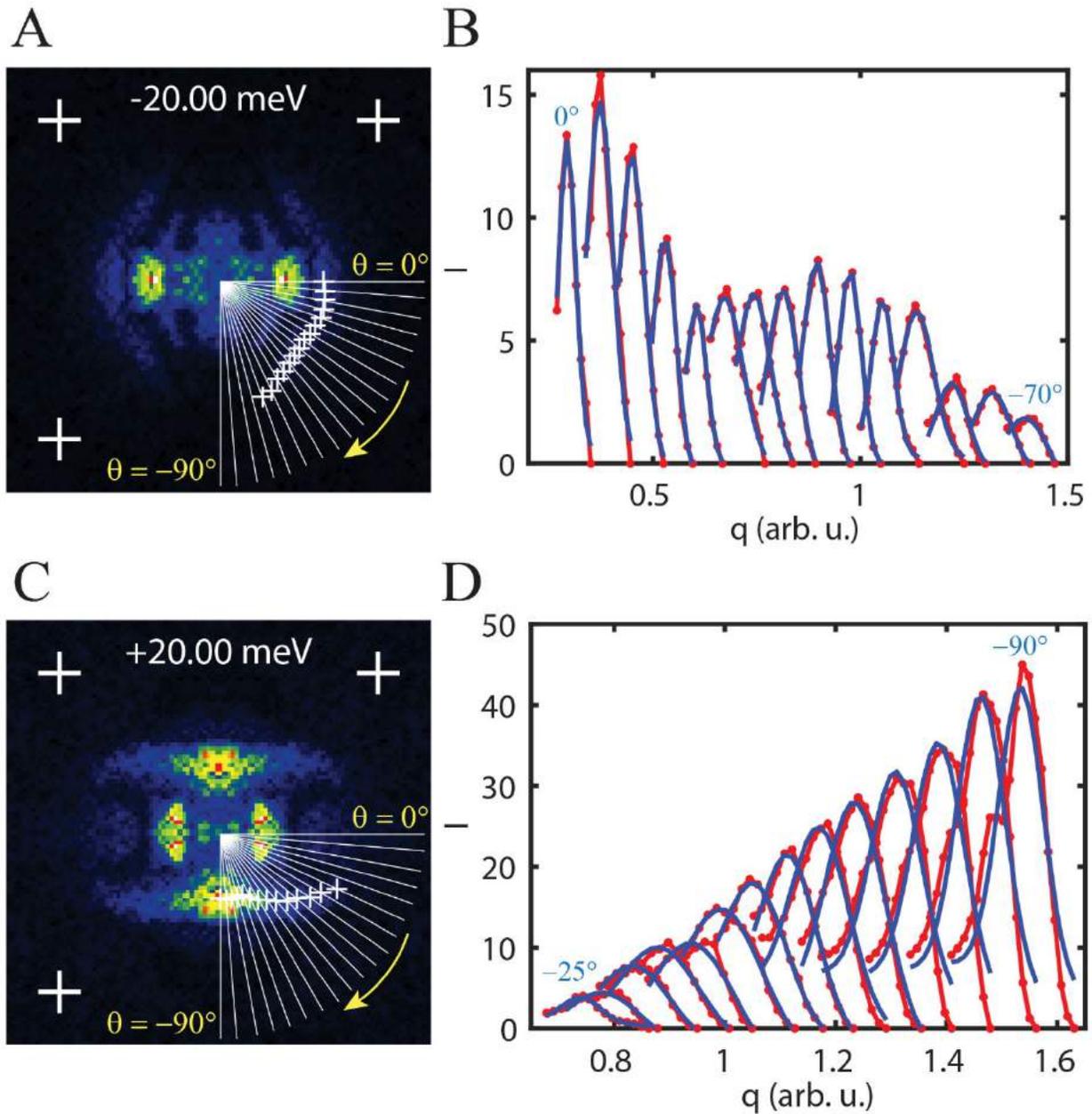

**Figure S9| Angular dependence of QPI intensity for intraband scattering due to the $\alpha$- and $\varepsilon$-pocket. A,** Measured $|L(\mathbf{q}, -20\ \text{meV})|$; white lines mark the angular cuts, and white crosses mark the extracted peak location. **B,** Angular dependence of peaks due to intraband scattering of $\alpha$-pocket at -20 meV in QPI line-cuts and fitted Gaussian peaks. **C,** Measured $|L(\mathbf{q}, +20\ \text{meV})|$; white lines mark the angular cuts, and white crosses mark the extracted peak location. **D,** Angular dependence of peaks due to intraband scattering of $\varepsilon$-pocket at +20 meV in QPI line-cuts and fitted Gaussian peaks. In B and D, both peaks and Gaussian curves have been subtracted by the minimum value of each peak's intensity so that all curves start at 0. Additionally, peaks and Gaussian curves have been shifted horizontally in q to enhance visibility.

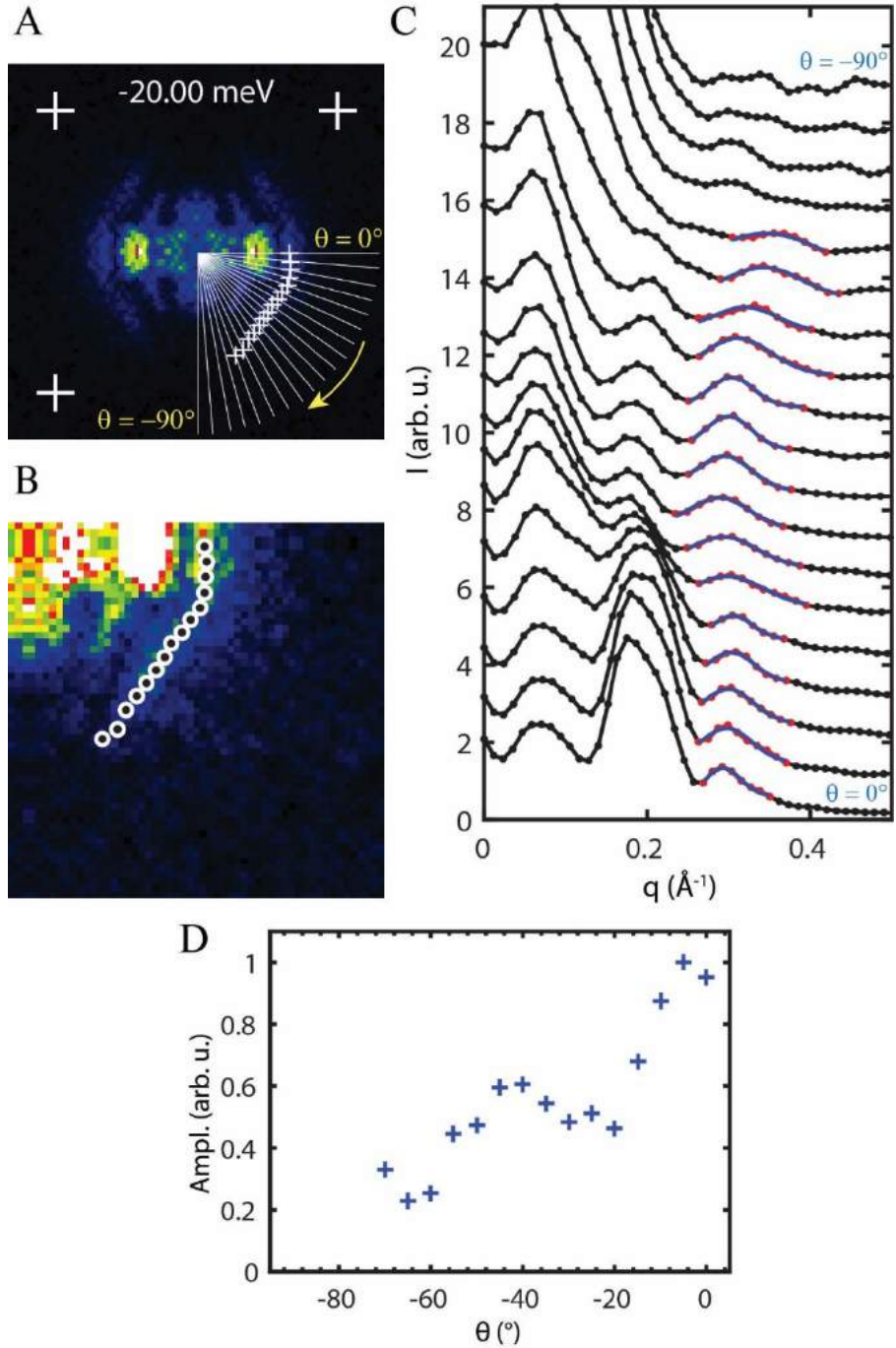

**Figure S10| Angular dependence of QPI intensity for intraband scattering due to the $\alpha$-pocket at -20 meV. A,** Measured $|L(\mathbf{q}, -20\ \text{meV})|$; white lines mark the angular cuts, and white crosses mark the extracted peak location from line-cuts in C. **B,** Magnification of lower right quarter of A. Black dots with white circles mark the extracted peak location from line-cuts in C. **C,** Angular line-cuts through QPI. Line-cuts have been scaled and shifted vertically to increase visibility. Blue curves represent the fit consisting of a Gaussian with constant and linear background. **D,** Amplitudes of the Gaussian fit as a function of angle.

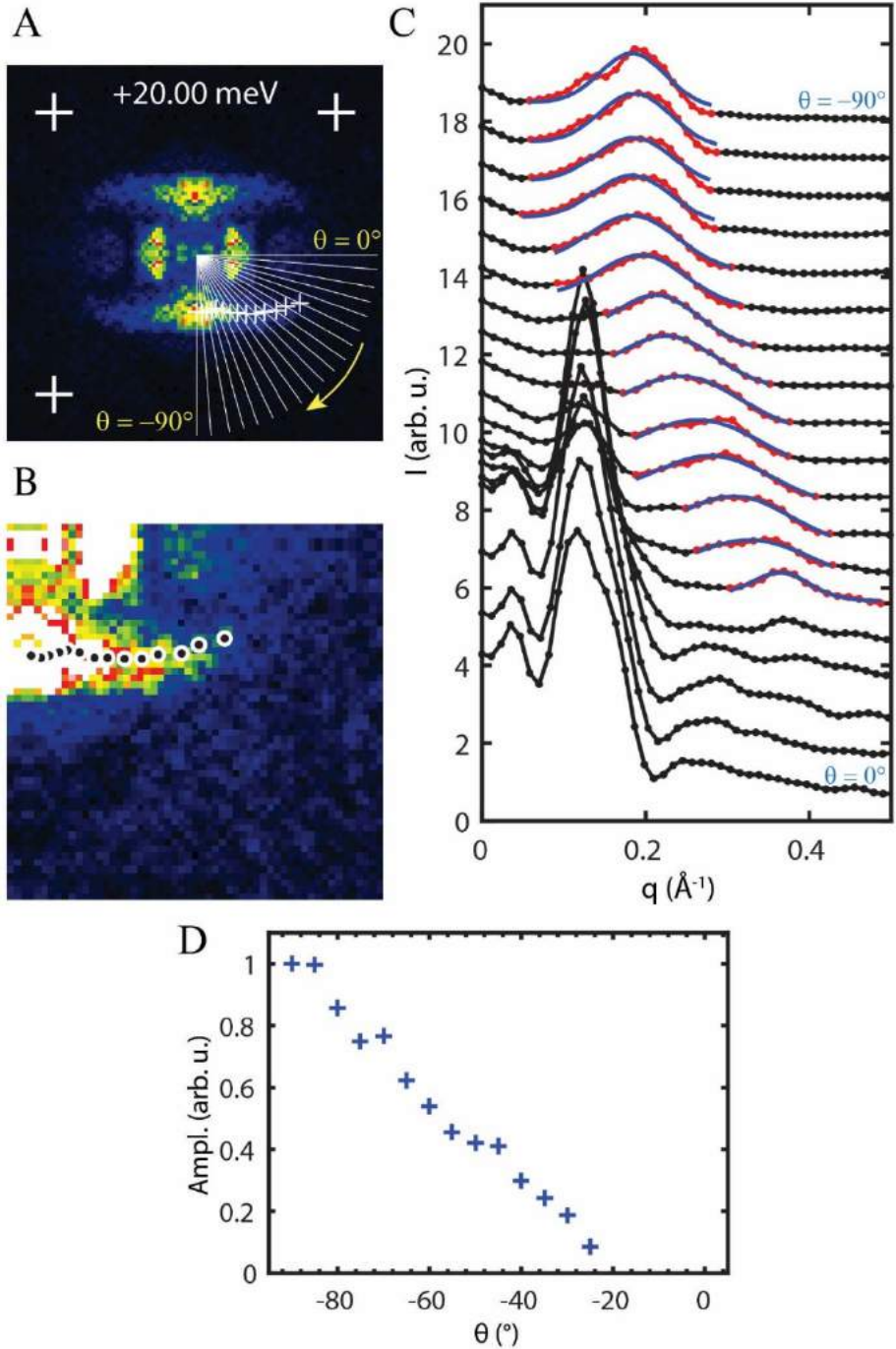

**Figure S11| Angular dependence of QPI intensity for intraband scattering due to the $\varepsilon$-pocket at +20 meV. A,** Measured $|L(\mathbf{q}, +20\text{ meV})|$; white lines mark the angular cuts, and white crosses mark the extracted peak location from line-cuts in C. **B,** Magnification of lower right quarter of A. Black dots with white circles mark the extracted peak location from line-cuts in C. **C,** Angular line-cuts through QPI. Line-cuts have been scaled and shifted vertically to increase visibility. Blue curves represent the fit consisting of a Gaussian with constant and linear background. **D,** Amplitudes of the Gaussian fit as a function of angle.

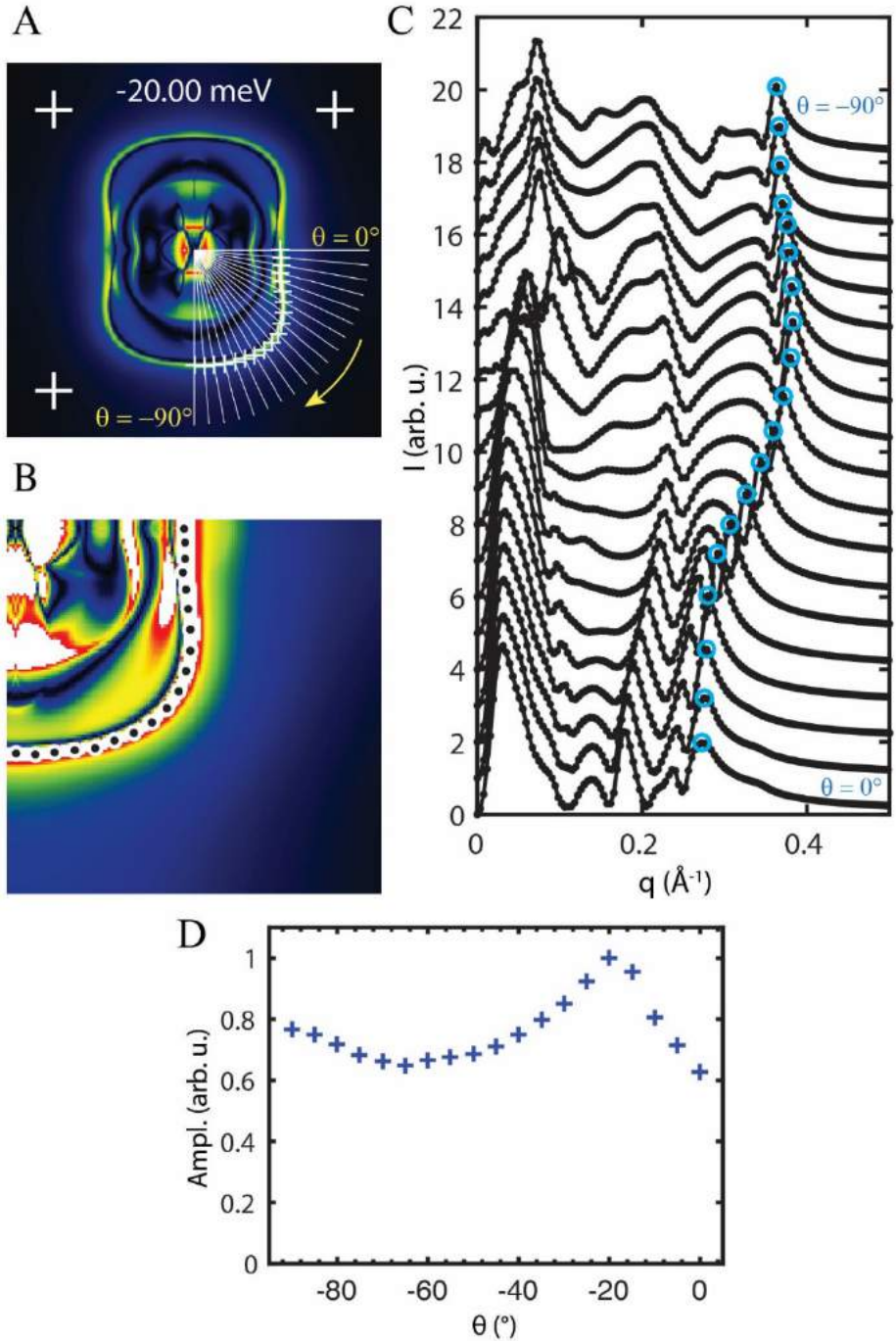

**Figure S12| Simulated angular dependence of QPI intensity for intraband scattering due to the $\alpha$-pocket at -20 meV for the case of equal quasiparticle weights. A,** Simulated |L(**q**, -20 meV)|; white lines mark the angular cuts, and white crosses mark the extracted peak location from line-cuts in C. **B,** Magnification of lower right quarter of A. Black dots with white circles mark the extracted peak location from line-cuts in C. **C,** Angular line-cuts through QPI. Line-cuts have been scaled and shifted vertically to increase visibility. Blue circles mark the peak location. **D,** Peak amplitudes as a function of angle.

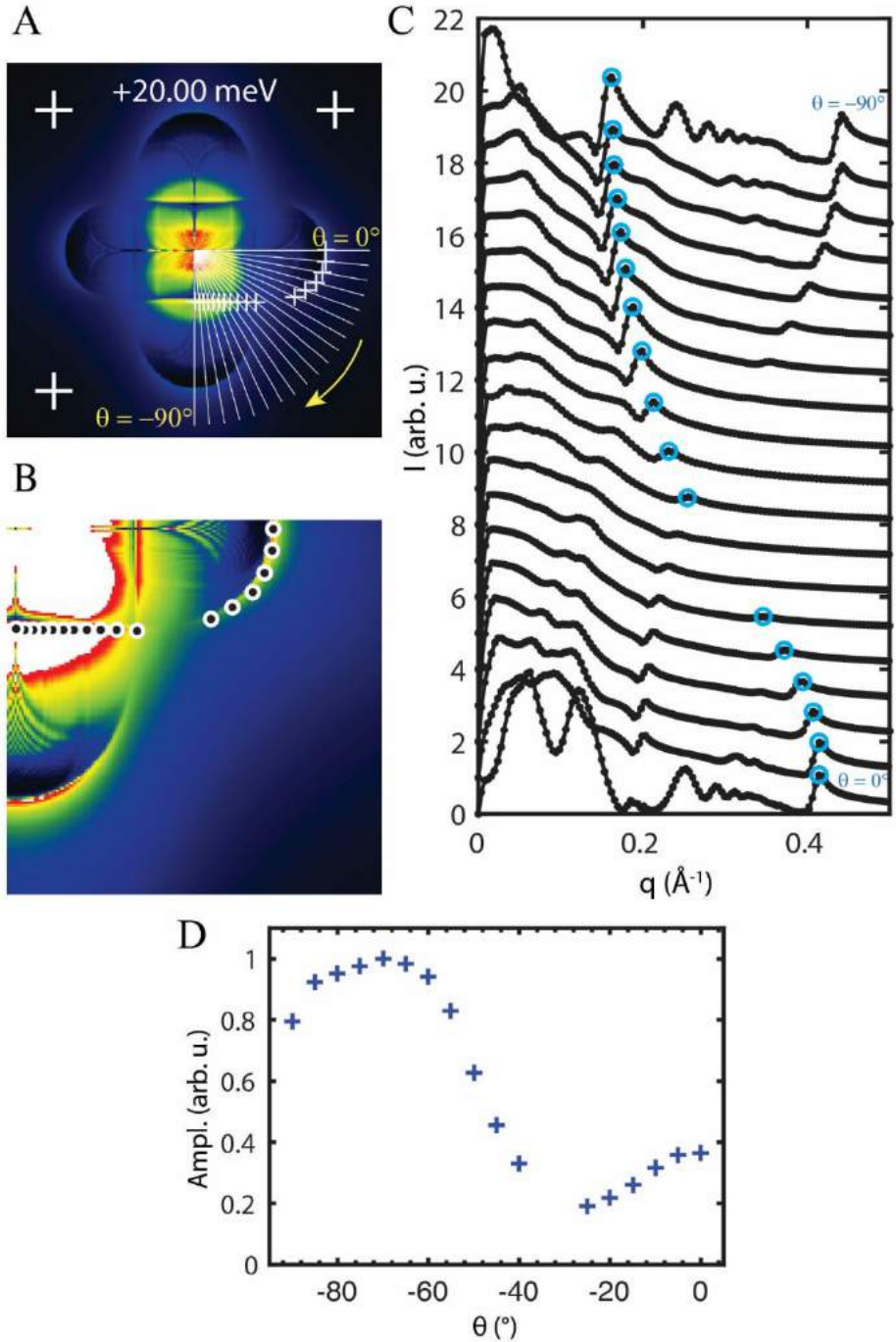

**Figure S13| Simulated angular dependence of QPI intensity for intraband scattering due to the $\varepsilon$-pocket at +20 meV for the case of equal quasiparticle weights. A,** Simulated $|L(\mathbf{q}, +20$ meV$)|$; white lines mark the angular cuts, and white crosses mark the extracted peak location from line-cuts in C. **B,** Magnification of lower right quarter of A. Black dots with white circles mark the extracted peak location from line-cuts in C. **C,** Angular line-cuts through QPI. Line-cuts have been scaled and shifted vertically to increase visibility. Blue circles mark the peak location. **D,** Peak amplitudes as a function of angle.

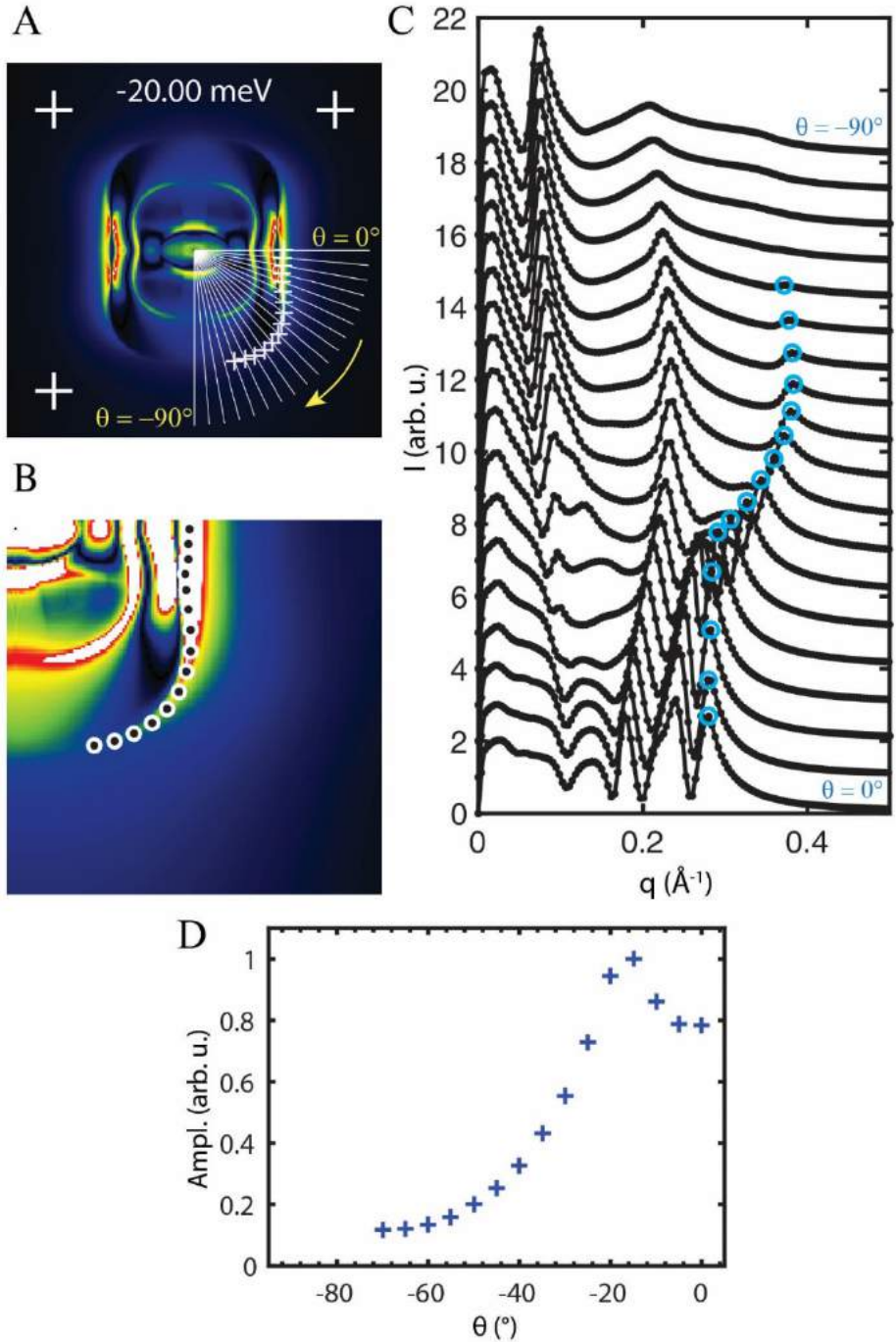

**Figure S14| Simulated angular dependence of QPI intensity for intraband scattering due to the $\alpha$-pocket at -20 meV for the case of orbital-selective quasiparticle weights. A,** Simulated $|L(\mathbf{q}, -20\ \text{meV})|$; white lines mark the angular cuts, and white crosses mark the extracted peak location from line-cuts in C. **B,** Magnification of lower right quarter of A. Black dots with white circles mark the extracted peak location from line-cuts in C. **C,** Angular line-cuts through QPI. Line-cuts have been scaled and shifted vertically to increase visibility. Blue circles mark the peak location. **D,** Peak amplitudes as a function of angle.

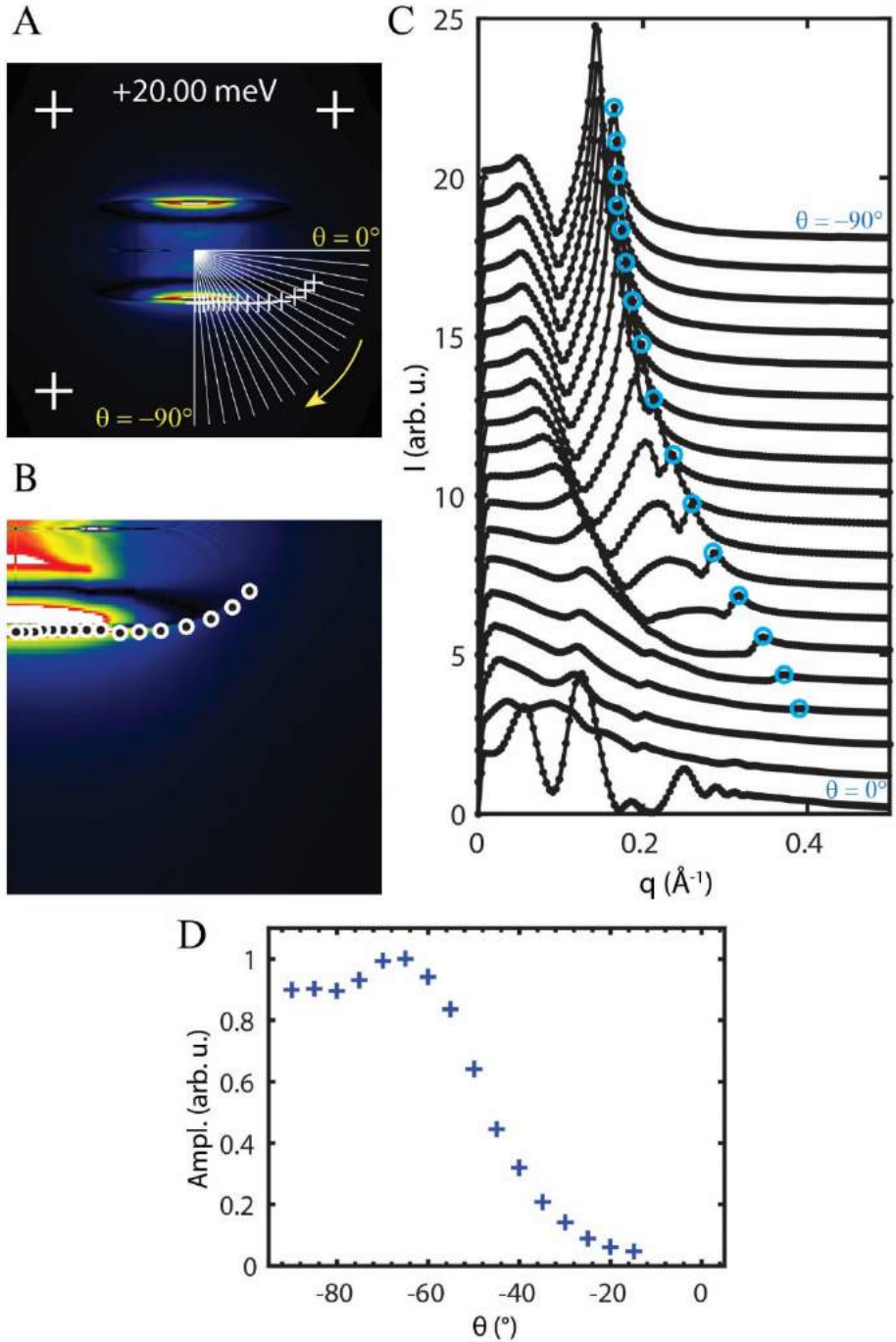

**Figure S15| Simulated angular dependence of QPI intensity for intraband scattering due to the $\varepsilon$-pocket at +20 meV for the case of orbital-selective quasiparticle weights.** **A,** Simulated $|L(\mathbf{q}, +20\text{ meV})|$; white lines mark the angular cuts, and white crosses mark the extracted peak location from line-cuts in C. **B,** Magnification of lower right quarter of A. Black dots with white circles mark the extracted peak location from line-cuts in C. **C,** Angular line-cuts through QPI. Line-cuts have been scaled and shifted vertically to increase visibility. Blue circles mark the peak location. **D,** Peak amplitudes as a function of angle.

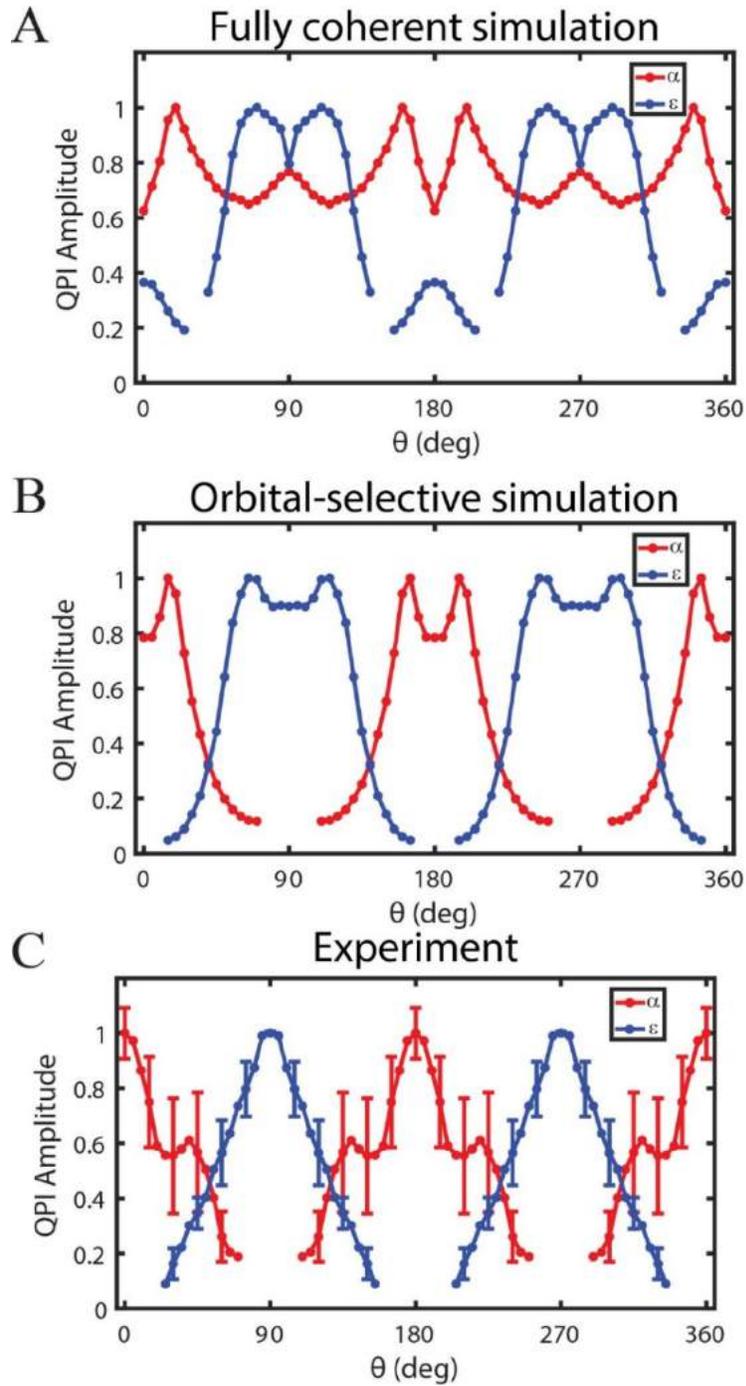

**Figure S16| Comparison of simulated and measured angular dependence of QPI intensity for the $\alpha$- and $\varepsilon$-pocket. A,** Simulated QPI amplitude as a function of angle for the case of all equal quasiparticle weights Z=1. **B,** Simulated QPI amplitude as a function of angle for the case of orbital-selective quasiparticle weights. **C,** Measured QPI amplitude as a function of angle.

Figure S17 shows a comparison of the simulated scattering of a single defect in real space to the measured $L(r, E)$. There is a stark contrast between simulation and experiment: The modeled scattering exhibits significantly more periods of oscillations than experiment. At this point, the precise cause of this effect is unclear. As experiment always has multiple scatterers present there could be some interference effect between them far from the individual scattering centers. Also, strong correlations in FeSe could lead to overall decoherence of quasiparticles, where some orbitals are more strongly affected than others. QPI is sensitive to differences in quasiparticle weight ratios rather than differences of absolute quasiparticle weight numbers.

However, no matter the cause we can investigate possible effects on our analysis of the angular dependence of QPI intensity. For this we fit a line-cut through a single defect to a constant plus a damped sinusoid: $c_0 + A_1 \exp\left[-\frac{(r-r_P)^2}{2\sigma^2}\right](A_2 \cos(qr + \phi))$. We choose 20 meV for the fit as it has very strong signal, and should only contain scattering along $q_y$ from the $\varepsilon$-pocket. To test if the extracted Gaussian is consistent with the suppression along $q_x$ as well as scattering at energies below the chemical potential we plot the Gaussian next to the line-cuts through the single defect for these scenarios. The comparison demonstrates that the suppression is virtually identical, both below and above the chemical potential and also for the two perpendicular directions. This is additional strong evidence that the scatterer is not anisotropic.

We use this Gaussian to modify the simulation. In real space, this is a multiplication with the Gaussian centered at the defect center. In Fourier space, the corresponding operation is convolution with the Fourier transform of the Gaussian. We then repeat the amplitude extraction from the line-cuts. Due to the convolution, there is now stronger background signal, and the amplitude is determined by taking the peak value and subtract from it the value at the end of the tail of the skewed curve. The extracted amplitudes still show the overall decrease in intensity as a function of angle. In contrast to the simulation which has not been convolved there is no more fine structure due to nesting. This could explain why the nesting structure is not observed in experiment. The overall picture remains the same. Orbital-selective decoherence leads to a strong decrease of QPI intensity for intraband scattering of the $\alpha$- and $\varepsilon$-pocket.

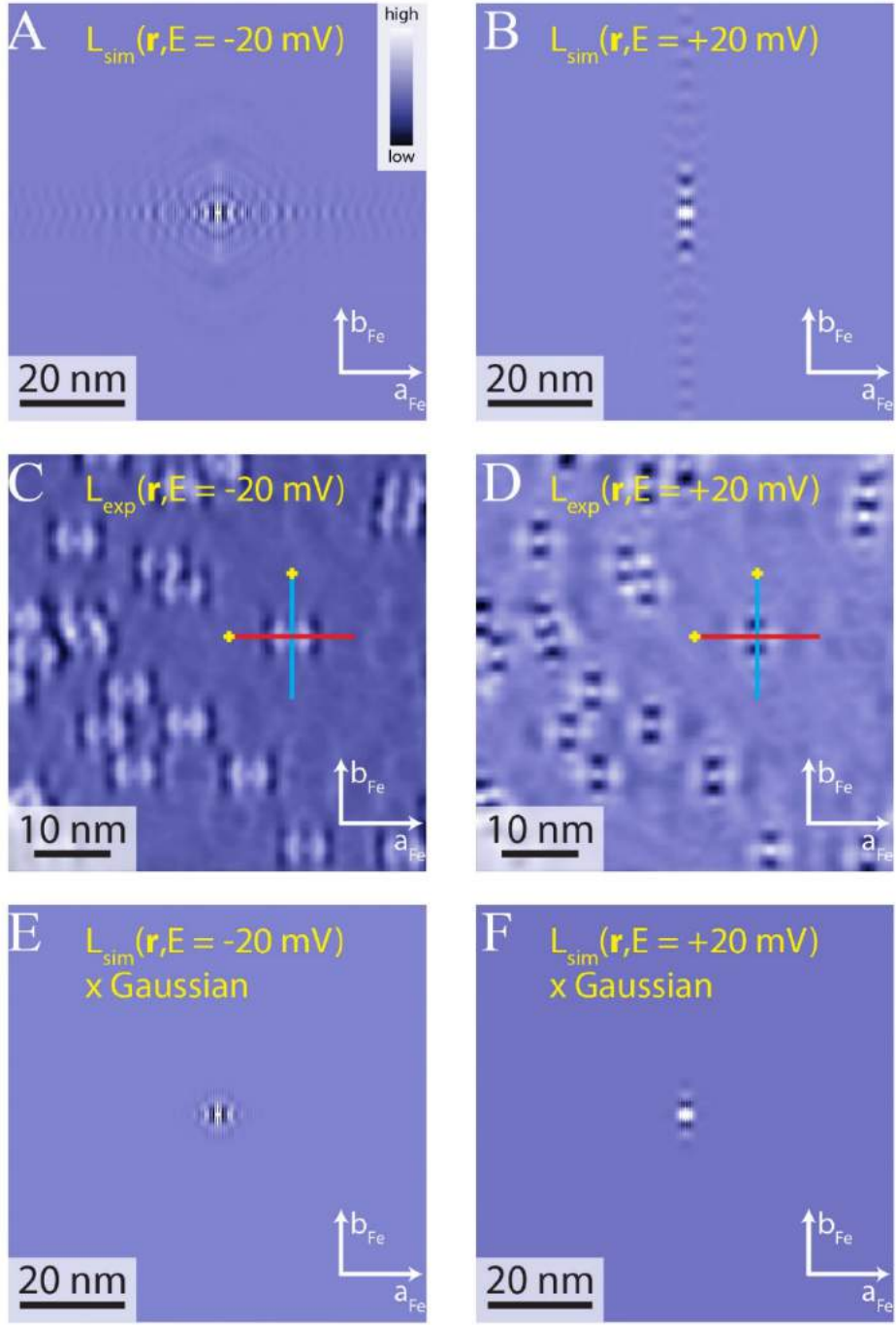

**Figure S17| Simulate the observed increased decoherence in experiment. A,** Simulated single scatterer at -20 meV. **B,** Simulated single scatterer at +20 meV. **C,** Measured QPI in real space at -20 meV. Red and blue lines represent line-cuts through isolated single impurity. **D,** Measured QPI in real space at +20 meV. Red and blue lines represent line-cuts through isolated single impurity. **E,** Simulated single scatterer at -20 meV multiplied with a Gaussian extracted from a fit to experiment, see Fig. S17. **F,** Simulated single scatterer at +20 meV multiplied with a Gaussian extracted from a fit to experiment, see Fig. S18.

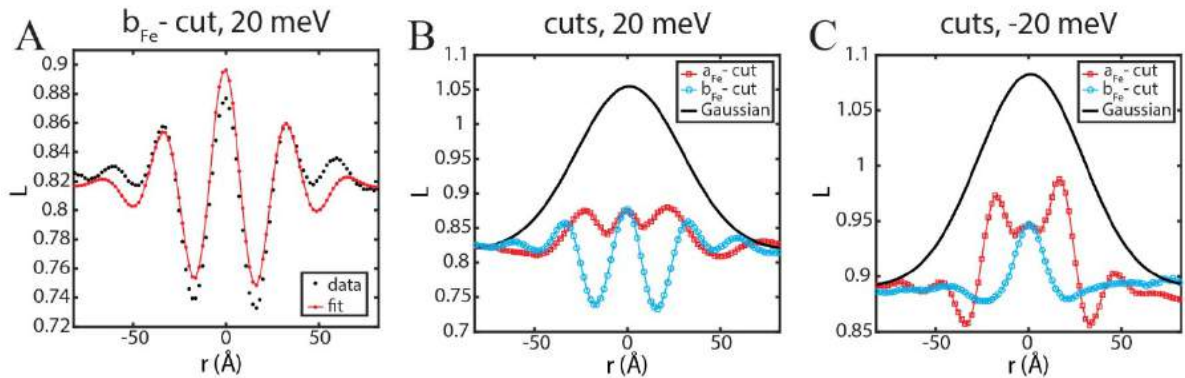

**Figure S18| Line-cuts through single defect quasiparticle interference in real space. A,** Line-cut along $b_{Fe}$ at +20 meV fitted using a constant and a damped sinusoid (specifically sinusoid multiplied by a Gaussian). **B,** Comparison of line-cuts through defect at +20 meV and Gaussian from fit in A. **C,** Comparison of line-cuts through defect at -20 meV and Gaussian from fit in A.

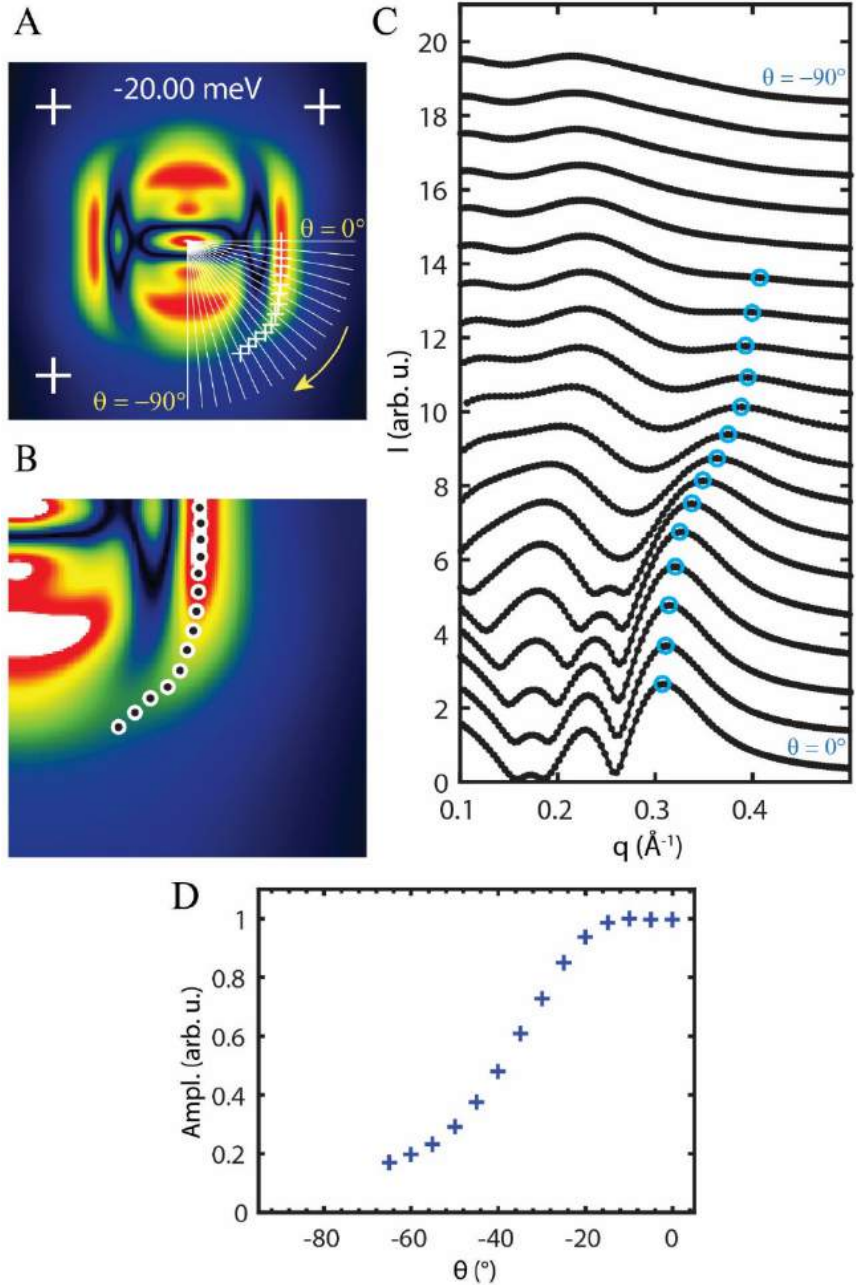

**Figure S19| Simulated angular dependence of QPI intensity for intraband scattering due to the $\alpha$-pocket at -20 meV for the case of orbital-selective quasiparticle weights with additional broadening in q-space. A,** Simulated $|L(\mathbf{q}, -20\text{ meV})|$ convolved with a 2-D Gaussian; white lines mark the angular cuts, and white crosses mark the extracted peak location from line-cuts in C. **B,** Magnification of lower right quarter of A. Black dots with white circles mark the extracted peak location from line-cuts in C. **C,** Angular line-cuts through QPI. Line-cuts have been scaled and shifted vertically to increase visibility. Blue circles mark the peak location. **D,** Peak amplitudes as a function of angle.

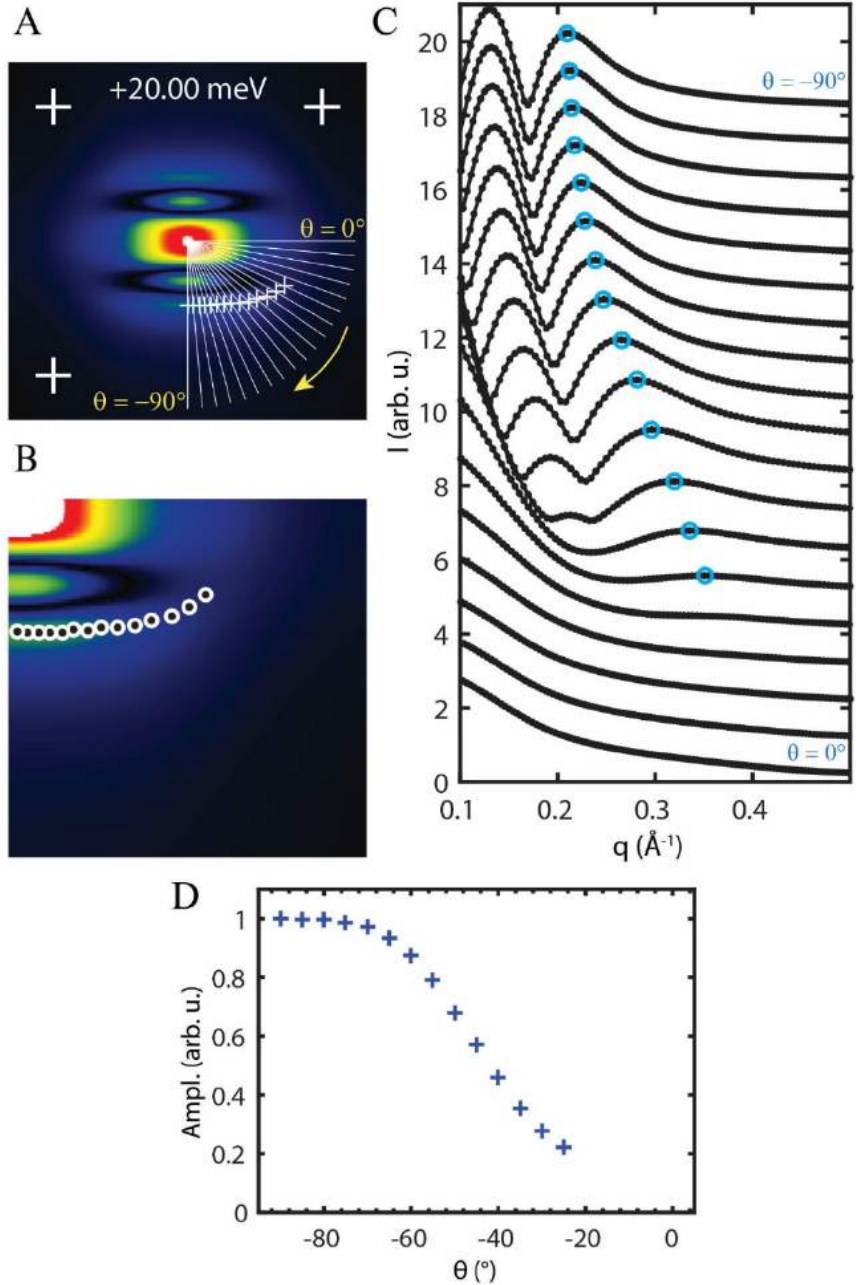

**Figure S20| Simulated angular dependence of QPI intensity for intraband scattering due to the $\varepsilon$-pocket at +20 meV for the case of orbital-selective quasiparticle weights with additional broadening in q-space.** **A,** Simulated $|L(\mathbf{q}, +20\text{ meV})|$ convolved with a 2-D Gaussian; white lines mark the angular cuts, and white crosses mark the extracted peak location from line-cuts in C. **B,** Magnification of lower right quarter of A. Black dots with white circles mark the extracted peak location from line-cuts in C. **C,** Angular line-cuts through QPI. Line-cuts have been scaled and shifted vertically to increase visibility. Blue circles mark the peak location. **D,** Peak amplitudes as a function of angle.

## 7. QPI measurements above and below superconducting $T_C$ using G and L

In order to reduce thermal smearing, we record QPI for energies |E| > 8.75 meV at 4.2 K. Before deciding to measure at 4.2 K we confirmed that the QPI signatures and E(*k*) dispersions of quasiparticles |E|>8.75meV which we report, are indistinguishable below and above Tc. The new data movie M4 which we have added to the SM, as well as one figure with selected energies from it (Figure S21), demonstrate these points.

Additionally, Fig. S21 and movie M4 demonstrate directly that G and L measurements detect the same anisotropic QPI phenomena as reported throughout this work.

## 8. Discussion of possible asymmetric scatterer

If the structure factor of the scattering potential was C2 symmetric along one axis of the crystal, it could be expected that this asymmetry should influence the QPI over a wide range of energies above and below the chemical potential. The energy dependence of the anisotropy contradicts such a simple symmetry argument. Movie M5 shows a comparison of theoretically predicted QPI with and without orbital-selective Z in real space and measured QPI for a single defect. For these two choices, only the simulation with orbital-selective Z reproduces the strong anisotropy observed in experiment. Furthermore, the orbital-selective simulation naturally contains the energy dependent rotation of the anisotropic scattering that is such a striking characteristic of FeSe. A theory with fully coherent quasiparticles Z=1 throughout, but an anisotropic scatterer would never be able to account for the observed energy dependence of anisotropic scattering.

Additional evidence for the orbital-selective quasiparticle weight origin of the anisotropic scattering comes from defects located within twin boundaries between orthorhombic domains. Here we concentrate on the scattering along the major Fe directions. The result is shown Fig. S22. If the scatterer itself was the origin of the anisotropic scattering one would expect that the C2-symmetric scattering pattern might change as a function of proximity to the twin boundary, and eventually become C4 symmetric therein, which is evidently not the case. What actually happens is that scattering of states only with $d_{zy}$ orbital content occurs from both sides of the TB. This indicates that it is orbital selectivity that is the microscopic cause of the intensely anisotropic QPI in FeSe.

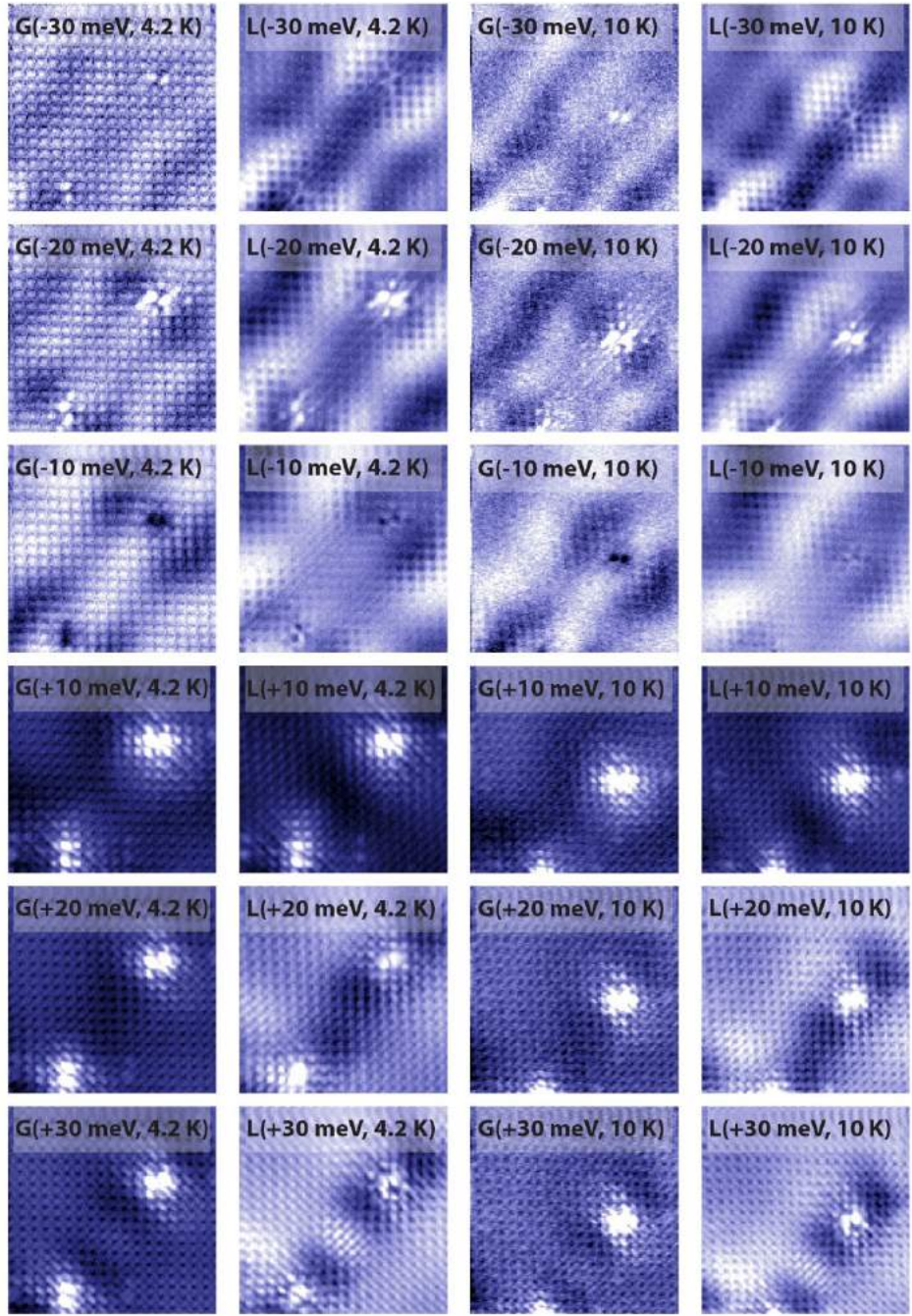

**Figure S21| Measured QPI in real space above and below $T_C$.** Measured G and L images for select energies below (columns 1,2) and above $T_C$ (columns 3,4). Movie M4 contains all these data and is a good example of how the anisotropic QPI signal from which we derive E(**k**) is indeed independent of temperature and equally present in G and L maps.

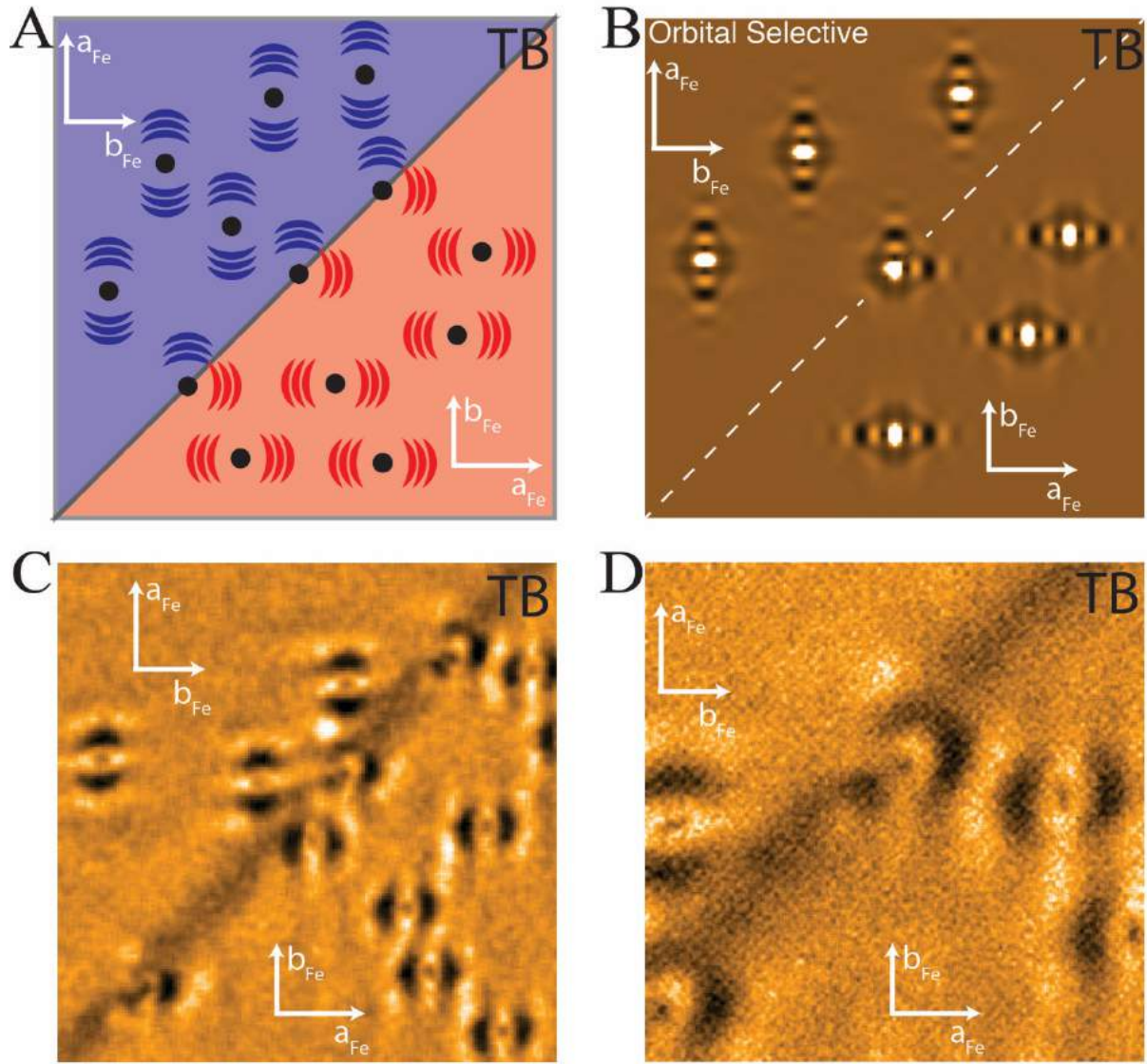

**Figure S22| Scattering from defects in very close proximity to a twin boundary. A,** In the orbital-selective Z scenario the scattering anisotropy originates from the distinct quasiparticle weights of the Fe d-electrons. As a consequence, scatterers located between two orthorhombic domains are expected to be reduced from C2 to C1 symmetry, and to a good approximation should equal the sum of scattering in the two domains. **B,** Simple simulation based on combining two diagonal halves for defects at the twin boundary. **C,** Measured scattering for several defects in close and very close proximity to a twin boundary. **D,** Magnified view of scattering by one of the defects very close to the twin boundary.

**Movie M1**

Movie contains predicted simulations of quasiparticle interference pattern assuming a fully coherent band structure. The simulations are computed within T matrix approach. The corners of the FOV in the Fourier space mark $\left(\pm \frac{\pi}{2a_{Fe}}, \pm \frac{\pi}{2b_{Fe}}\right)$ points. The energy range for DOS images is -25 meV to +25 meV, with 1.25 meV spacing.

**Movie M2**

Movie contains predicted simulations of quasiparticle interference pattern assuming orbitally selective quasiparticle weights. The simulations are computed within T matrix approach. The corners of the FOV in the Fourier space mark $\left(\pm \frac{\pi}{2a_{Fe}}, \pm \frac{\pi}{2b_{Fe}}\right)$ points. The energy range for DOS images is -25 meV to +25 meV with 1.25 meV spacing.

**Movie M3**

Movie contains measured quasiparticle interference pattern within -25 meV to +25 meV energy range acquired with 1.25 meV spacing.

**Movie M4**

Measured quasiparticle interference pattern in real space within -35 meV to +35 meV energy range acquired with 1 meV spacing at 4.2K (< $T_C$) and 10K (> $T_C$). Upper left corner G(**r**, eV, 4.2K). Upper right corner G(**r**, eV, 10K). Lower left corner L(**r**, eV, 4.2K). Lower right corner L(**r**, eV, 10K).

**Movie M5**

Movie contains real space simulated and measured quasiparticle interference pattern within -25 meV to +25 meV energy range acquired with 1.25 meV spacing. From left to right: fully coherent simulation with all Z=1, experiment, and simulation with orbital-selective Z as used throughout this work. For the experiment, short wavelength modulations, as for example the atomic lattice, have been filtered out in order to make comparison of measured and simulated QPI easier.